\documentstyle[aps,pre,preprint]{revtex}
\begin{document}
\preprint{SNUTP 93-66}
\draft
\title{Interacting Domain Walls and the Five--Vertex Model}
\author{Jae Dong Noh and Doochul Kim}
\address{ Department of Physics and Center for Theoretical Physics,
Seoul National University,  Seoul 151--742, Korea}
%\date{\today}
\maketitle

\begin{abstract}
We investigate the thermodynamic and
critical properties of an interacting domain wall model which is
derived from the triangular lattice antiferromagnetic Ising model
with the anisotropic nearest and next nearest neighbor interactions.
The model is equivalent to the general five--vertex model.
Diagonalizing the transfer matrix exactly by the
Bethe Ansatz method, we obtain the phase diagram displaying the
commensurate and incommensurate (IC) phases separated by the
Pokrovsky--Talapov transitions.
The phase diagram exhibits commensurate phases where the
domain wall density $q$ is locked at the values of $0$, $1/2$ and
$1$.
The IC phase is a critical state described by the
Gaussian fixed point. The effective Gaussian coupling constant is
obtained analytically and numerically for the IC phase
using the finite size scaling  predictions of the conformal field
theory.  It takes the value $1/2$ in the non-interacting limit and
also at the boundaries of $q=0$ or $1$ phase and the value $2$ at the
boundary of $q=1/2$ phase, while it varies smoothly throughout the
IC region.
\end{abstract}
\pagebreak

\section{Introduction}\label{sec1}
There has been much interest in the two dimensional statistical
mechanical systems which exhibit modulated phases on the periodic
substrate~\cite{Bak}.
Among those systems are monolayers of
physisorbed gas on solid surface which display
{\em incommensurate} (IC) and {\em commensurate} (C) phases.
In the domain wall description of IC phases~\cite{FraMer}, domain
walls separating commensurate patches are considered as the basic
fluctuating degrees of freedom.  The domain walls can be arranged
either parallel to each other (striped domain wall) or in
hexagonal pattern (honeycomb domain wall) depending on the domain
wall crossing energy~{\cite{CopFis}}.
The simplest type of commensurate--incommensurate (C--IC)
transition is the Pokrovsky--Talapov
(PT) transition~\cite{PokTal} which describes the transition into
striped IC phase. Here, the fluctuations of the striped domain wall
cause an effective repulsive interaction between walls. The
interaction  varies
as $1/l^3$ if $l$ is the average distance between walls.
Due to this repulsive interaction between domain walls, the C--IC
transition to the striped IC phase
is a continuous transition with the specific heat exponent
$\alpha=1/2$ and the domain wall density displays a square root
dependence on the chemical potential of domain wall if we
approach the phase boundary from the incommensurate side.
The theory is explicitly realized in fermion models of striped
IC phases where domain walls are represented as world lines of
fermions living in one dimensional chain.
Free fermion model is also obtained as  low temperature
approximation to the ANNNI model~\cite{VilBak}. In these models,
the IC phase is a critical phase where the correlation functions
decay by the power laws of the distance rather than by the
exponential function of the distance. Recently, Park and Widom
showed that the IC phase modeled by free fermion hamiltonian
is described in the continuum limit by the
Gaussian model with the coupling constant $g=1/2$ by explicit
calculation of the toroidal partition function~{\cite{ParWid}}.
Effect of domain wall interaction has also been studied in the
fermion model derived from an approximation to the
ANNNI model~\cite{GryCev} and in a phenomenological
model~\cite{ParWid}.

In this paper, we consider an exactly solvable {\em interacting}
domain wall model derived from the triangular lattice
antiferromagnetic Ising model (TAFIM). It is well known that the
TAFIM with the only nearest neighbor coupling has infinitely
degenerate ground states due to frustration on each elementary
triangles. Each ground state can be mapped into a configuration of
covering the plane by three types of diamonds. Bl\"{o}te and
Hilhorst~\cite{BloHil} introduced a solid-on-solid
model derived from these configurations.
Regarding two types of diamonds as domain wall
excitations, one also obtains a striped domain wall configuration.
Bl\"{o}te and Hilhorst~\cite{BloHil} utilized this connection to
obtain exact solution to the non-interacting domain  wall problem.
As the fugacities of walls change, there is a phase transition
from an ordered phase to the critically disordered phase which is
described by the Gaussian fixed point with the coupling constant
$g=2$. The nature of the transition is found to be that of the PT
transition~\cite{PokTal}. Nienhuis {\em et al.}~{\cite{NieHil}}
identified
various spin wave and vortex operators of the Gaussian model in
terms of the solid-on-solid model and argued that infinitesimal
next nearest neighbor (nnn) interactions and magnetic field in
TAFIM would change the coupling constant $g$ of the Gaussian model.
 From this they suggested a schematic phase diagram in the parameter
space composed of the nearest neighbor interactions, the nnn
interactions and the external magnetic field.
More recently the effect of the external magnetic field on $g$ has
been studied by Bl\"{o}te {\it et al.}~\cite{BloNWH} and the
behaviors predicted in Ref.~\cite{NieHil} is confirmed.

We show in Sec.~\ref{sec2} that the ground state configurations of
the TAFIM under the general boundary conditions are equivalent to
the striped domain wall configurations. When the nnn interactions
in the TAFIM are turned on in an anisotropic manner, they
correspond to extra energies between adjacent domain walls.
Only the same types of walls can touch each other
and there are two types of wall interactions.
We also show in Sec.~\ref{sec2} that the striped domain wall
configuration is exactly mapped to the arrow configuration of the
5--vertex model.
But, if both types of wall interactions are present, the
Boltzmann weight cannot be represented by a product of vertex
weights. However, when only one type of domain walls interacts each
other, it can be written as a product of vertex weights and the
partially interacting domain wall model reduces to the general
5--vertex model.

In Sec.~\ref{sec3}, we diagonalize
the transfer matrix of the 5-vertex model using the Bethe Ansatz
method. We develop  Bethe ansatz  solutions both for domain wall and
domain wall hole. From these solutions, we obtain full phase diagram
of the partially interacting model.
The phase diagram displays the C and IC phases separated
by the PT transition and the first order transition. It also
exhibits a new commensurate phase where the domain wall density is
locked to the value $1/2$ for a range of the chemical potential of the
wall. This phase does not appear in the non-interacting domain wall
models and is a feature resulting from the domain wall interactions.
This is akin to the antiferromagnetically ordered phase of the ANNNI
model.

In Sec.~\ref{sec4}, we investigate the critical properties of the IC
phase. It is shown that the interaction between domain walls causes
a continuous variation of the coupling constant $g$ of the Gaussian
model resulting in non-universal critical behaviors. It is studied
by analytic perturbative calculations
and numerical calculations. We discuss and summarize our result in
Sec.~\ref{sec5} and present discussions on the Yang--Baxter equation
and the calculation of the modular covariant partition functions
of the $T=0$ TAFIM under the general boundary conditions
in Appendix~\ref{appA} and \ref{appB}, respectively.

\section{Transfer Matrix Formulation of Interacting Domain Wall
Model}\label{sec2}
We write the hamiltonian ${\cal H}$ including $1/kT$ of the TAFIM
with the nearest and next nearest neighbor interaction as
\begin{equation}\label{hm}
{\cal H}=-\sum_{<ij>}(K+\delta_a)s_i s_j - \sum_{<\!<ij>\!>}\!\!\!'
\varepsilon_a s_i s_j - K {\cal N}
\end{equation}
where $s_i=\pm1$ is an Ising spin variable at site $i$, the first
(second) sum is over the nearest (next nearest) neighbor pairs of
sites, $K+\delta_a$ ($\varepsilon_a$), $a=1,2,3,$ are the
anisotropic nearest  (next nearest) neighbor couplings whose index
$a$ depends on the direction of the bond $<\!ij\!>$ ($<\!<\!ij\!>\!>$)
as shown in Fig.~1(a), and finally
${\cal N}$ is the number of lattice sites.

Monte Carlo simulation and other studies~\cite{Lan,Hem} show that
this system has rich critical phenomena in the full parameter space.
But, we will only consider the zero temperature limit of this system.
By the zero temperature limit, we actually mean the infinite
coupling limit $ K\rightarrow -\infty$ leaving $\delta_j$'s and
$\varepsilon_j$'s finite. Eq.~(\ref{hm}) in this limit will be called
the $T=0$ TAFIM.
Here, only those configurations which have precisely one pair of
parallel spins around each elementary triangle are energetically
allowed.  Though this imposes much restriction on the spin
configurations, it is important to study
this limiting case because the $T=0$ TAFIM is equivalent to many
interesting problems, e.g. diamond and/or dimer covering
problem~\cite{BloHil} and
triangular solid-on-solid model~\cite{NieHil,ForeT}. Moreover,
the $T\neq 0$ behavior of the TAFIM can be inferred from the
$T=0$ behavior.

Here, we will show that the $T=0$ TAFIM with nnn interaction is
equivalent to the interacting striped domain
wall model where the nnn interaction $\varepsilon_j$ $(j=1,2)$
plays the role of  wall--wall interactions.
If we draw lines between all nearest
neighbor pairs of antiparallel spins for a given ground state
configuration of the TAFIM,
the resulting configuration is that of a covering of the plane
by diamonds.
Fig.~2 shows a typical TAFIM ground state and its corresponding
diamond covering configuration.
The three types of diamonds are called as type 1,2
and 3, respectively as shown in Fig.~1(b).
Strictly speaking, there is two-to-one correspondence
because of the global spin reversal symmetry of the
TAFIM in the absence of magnetic field.

 From a diamond covering configuration, a striped domain wall
configuration  is obtained
by regarding the diamonds of type 1 and 2 as domain wall excitations.
Type 3 diamonds are regarded as the vacuum. Thick lines on the faces
of type 1 and 2 diamonds in Fig~1(b) and Fig.~2 visualize the domain
walls.  A section of domain walls which is obtained from the diamond
of type 1 and 2 will be called the domain wall of type 1 and 2,
respectively. Two walls are defined to be interacting when their
sides touch each other. The thermodynamic parameters which control
the equilibrium property of interacting striped domain wall
system are the fugacities $x_1$ and $x_2$ of domain walls of type 1
and 2, respectively, and the fugacities $y_1$ and $y_2$ for each pair
of adjacent domain walls of type 1 and 2, respectively. Note that
different types of domain walls cannot be adjacent. The partition
function
${\cal Z}_{\mbox{{\tiny d.w.}}}$ for the interacting domain wall
model is
\begin{equation}\label{e.zdw}
{\cal Z}_{\mbox{{\tiny d.w.}}} = \sum x_1^{n_1} x_2^{n_2}
y_1^{l_1} y_2^{l_2}
\end{equation}
where the summation is taken over all striped domain wall
configurations and $n_i$ is the total length of domain wall of type
$i$ and $l_i$ is the total number of incidents where domain walls
of type $i$ touch each other and share a side, {\it i.e.} the number
of wall--wall interactions of type $i$.

When $\varepsilon_3=0$ in Eq.~(\ref{hm}), the energy of the $T=0$
TAFIM can be written in terms of $n_i$ and $l_i$. Nearest neighbor
interactions contribute~\cite{BloHil} simply
$$
\sum_{i=1,2,3} (-\delta_i+\delta_j+\delta_k) n_i
$$
where $(i,j,k)$ is the cyclic permutation of $(1,2,3)$ and $n_3=
{\cal  N} -n_1-n_2$ is the total number of type 3 diamonds.
 From now on we set $\delta_3=0$ without loss of generality.
To relate the nnn interaction energies to $l_i$, consider first
the bonds connecting nnn pair of sites along the direction 1. (See
Fig.~1(a).) They cross either (a) two type 2 domain walls or (b)
two type 3 diamonds or (c) one type 2 wall and one type 3 diamond
or (d) one type 1 domain wall. These possibilities are shown in
Fig.~3. If we let $n_a,n_b,n_c$ and $n_d$ be the number of cases
(a), (b), (c) and (d), respectively, the bonds contribute
$\varepsilon_1(n_a+n_b-n_c-n_d)$ to the energy. But one can easily
identify $n_d=n_1$ and $n_a=l_2$.
Moreover each of type 2 wall is crossed by two nnn bonds so that
it appears twice in the list of Fig.~3 while the total number of
type 2 walls counted in Fig.~3 is $2n_a+n_c$. Thus $2n_2=2n_a+n_c$.
These relations, together
with the sum rule $n_a+n_b+n_c+n_d={\cal N}$, give the energy
$$
\varepsilon_1 ( {\cal N} + 4 l_2 - 4 n_2 - 2 n_1 ) \ \ \ .
$$
Similar counting holds for nnn bonds along the direction 2.

Putting these together, the ground state energy of Eq.~(\ref{hm})
for $\varepsilon_3=0$ becomes
\begin{eqnarray}
E_0 &=&-2(\delta_1+\varepsilon_1+2\varepsilon_2)n_1
     -2(\delta_2+\varepsilon_2+2\varepsilon_1)n_2\nonumber \\
    && +4 \varepsilon_2 l_1 + 4 \varepsilon_1 l_2 +
    {\cal N} ( \delta_1 + \delta_2 + \varepsilon_1 +
\varepsilon_2 ) \ \ \ .
\end{eqnarray}
Thus the fugacities for the interacting domain wall model are
related to anisotropic coupling energies of the TAFIM model as
\begin{equation}
\begin{array}{ccl}
x_1&=&\exp\left[2(\delta_1-\delta_3)+2\varepsilon_1+
4\varepsilon_2\right] \\[2mm]
x_2&=&\exp\left[2(\delta_2-\delta_3)+2\varepsilon_2+
4\varepsilon_1\right] \\[2mm]
y_1&=&\exp\left[-4\varepsilon_2\right]\\[2mm]
y_2&=&\exp\left[-4\varepsilon_1\right]
\end{array}
\end{equation}

Next, we show that to each striped domain wall configuration, one
can assign a vertex configuration. To do this we deform the
triangular lattice into the square one as shown in Fig.~4.
One then finds that there are 5 types of unit squares. Fig.~5 shows
them together with assignment of vertex
configurations. If one works under the ice rule, the assignment
of vertices shown in Fig.~5 is unique modulo the arrow reversal.

In this way, we obtain one-to-one correspondence between striped
domain wall configurations and bond arrow configurations satisfying
the ice rule.  Vertical up arrow indicates the presence of a domain
wall. In the TAFIM language, vertical up arrows correspond to
horizontal nearest neighbor spin pairs which have opposite signs
and right arrows correspond to vertical nearest neighbor spin pairs
which have the same signs. The absence of the third vertex (or the
fourth upon arrow reversal) is a result of our deforming the
triangular lattice in the manner shown in Fig.~4. If it were
deformed in the opposite direction, it is the first vertex (or the
second upon the arrow reversal) which
does not appear. In any case, one obtains the 5--vertex
model configurations.

The 5--vertex model on the square lattice is obtained from the
6--vertex model by suppressing one of the first four vertices.
The 5--vertex model with special choice of
its vertex weights was first considered by Wu~\cite{LieWu} as a
limiting case of the 6--vertex model and is studied in connection
with the non-intersecting directed random walk~\cite{Bha} and the
directed percolation problem in three dimension~\cite{Wup}.
Recently, Gul\'{a}csi {\it et al.}~\cite{GulBL} studied its phase
diagram for a special case. The general 5--vertex
model is obtained by assigning arbitrary vertex weights to each
type of vertices but there are only three independent parameters
since the vertices 5 and 6 always occur in pairs along a row under
the periodic boundary condition and a global rescaling of
weights introduces only a trivial factor.

The partition function ${\cal Z}_{\mbox{{\tiny 5-v}}}$ of the
5--vertex model is
\begin{equation}\label{e.z5v}
{\cal Z}_{\mbox{{\tiny 5-v}}} = \sum w_1^{N_1} w_2^{N_2} w_4^{N_4}
w_5^{N_5} w_6^{N_6}
\end{equation}
where the summation is taken over all arrow configurations and $N_i$
is the number of the $i$-th vertex appearing in an arrow
configuration. Unfortunately, if we assign Boltzmann weight of an
arrow configuration as a product of local vertex weights, we can not
treat the fully interacting domain wall model ($\varepsilon_1\neq 0,
\varepsilon_2\neq 0$ case). However, if we restrict ourselves to the
special case $\varepsilon_1=0$ $(y_2=1)$, then Eq.~(\ref{e.zdw}) can
be expressed in the form of Eq.~(\ref{e.z5v}).
 From now on, we consider the partially interacting model where only
domain walls of type 1 interact.

The vertex of type 1 represents the case where the two domain walls
of type 1 are interacting. So, if we choose vertex weights as
\begin{equation}\label{e.vw}
\left\{
\begin{array}{ccl}
w_1 &=& x_1 y_1 \\
w_2 &=& 1 \\
w_4 &=& x_2 \\
w_5 = w_6 &=& \sqrt{x_1}
\end{array} \right.
\end{equation}
the partition function ${\cal Z}_{\mbox{{\tiny 5-v}}}$ becomes the
same as that of the partially interacting domain wall system;
\begin{equation}
{\cal Z}_{\mbox{{\tiny 5-v}}}=\sum x_1^{n_1}x_2^{n_2}y_1^{l_1} \ \ \ .
\end{equation}

We study the 5--vertex model using the transfer matrix.
Suppose the lattice has $M$ rows and $N$ columns, and
periodic boundary conditions are imposed in both directions.
Let ${\hbox{\boldmath $\alpha$}}=(\alpha_1,\cdots,\alpha_N)$ denote
the state of vertical arrows of one row. Then, as usual, we can
write the partition function ${\cal Z}_{\mbox{{\tiny 5-v}}}$ as
\begin{equation}\label{e.t5v0}
{\cal Z}_{\mbox{{\tiny 5-v}}} = \mbox{Tr}\ \  {\bf T}_{5-v}^M
\end{equation}
where ${\bf T}_{\mbox{{\tiny 5-v}}}$ is the $2^N$ by $2^N$ transfer
matrix with elements
\begin{equation}\label{TM}
{\bf T}_{\mbox{{\tiny 5-v}}} ({\hbox{\boldmath $\alpha$}},
{\hbox{\boldmath $\beta$}})=\sum_{\{\mu_i = \pm 1\}} \prod_{i=1}^N
W(\mu_i,\alpha_i | \beta_i,\mu_{i+1} ) \ \ \ .
\end{equation}
In Eq.~(\ref{TM}), $W(\mu,\alpha|\beta,\nu)$ is the weight of the
vertex configuration in the standard notation~\cite{Bax}.
Let ${\bf T}_L\ ({\bf T}_R)$ be the transfer matrix of the
5--vertex model with the first horizontal arrow fixed to the left
(right). This can be written graphically as

\begin{picture}(400,100)(-200,-45)
\thicklines

\put(-100,0){\line(1,0){100}}
 \put(-100,-0.5){\line(1,0){100}}
\put(12,-5){\makebox(0,0){,}}
\multiput(-80,20)(20,0){2}{\line(0,-1){40}}
 \multiput(-80.5,20)(20,0){2}{\line(0,-1){40}}
\put(-20,20){\line(0,-1){40}}
 \put(-20.5,20){\line(0,-1){40}}

\put(-80,-30){\makebox(0,0){$\alpha_1$}}
\put(-60,-30){\makebox(0,0){$\alpha_2$}}
\put(-20,-30){\makebox(0,0){$\alpha_N$}}
\multiput(-45,30)(5,0){3}{\circle*{1}}
\put(-80,30){\makebox(0,0){$\beta_1$}}
\put(-60,30){\makebox(0,0){$\beta_2$}}
\put(-20,30){\makebox(0,0){$\beta_N$}}
\multiput(-45,-30)(5,0){3}{\circle*{1}}

\multiput(-7,0)(-83,0){2}{\vector(-1,0){5}}

\put(-150,0){\makebox(0,0){${\bf{T}}_{L}({\hbox{\boldmath $\alpha$}}
\mid {\hbox{\boldmath $\beta$}}) \; = $}}

\put(115,0){\line(1,0){100}}
 \put(115,-0.5){\line(1,0){100}}
\multiput(135,20)(20,0){2}{\line(0,-1){40}}
 \multiput(134.5,20)(20,0){2}{\line(0,-1){40}}
\put(195,20){\line(0,-1){40}}
 \put(194.5,20){\line(0,-1){40}}

\put(135,-30){\makebox(0,0){$\alpha_1$}}
\put(155,-30){\makebox(0,0){$\alpha_2$}}
\put(195,-30){\makebox(0,0){$\alpha_N$}}
\multiput(170,30)(5,0){3}{\circle*{1}}
\put(135,30){\makebox(0,0){$\beta_1$}}
\put(155,30){\makebox(0,0){$\beta_2$}}
\put(195,30){\makebox(0,0){$\beta_N$}}
\multiput(170,-30)(5,0){3}{\circle*{1}}

\multiput(205,0)(-83,0){2}{\vector(1,0){5}}

\put(65,0){\makebox(0,0){$\;{\bf{T}}_{R}({\hbox{\boldmath $\alpha$}}
\mid {\hbox{\boldmath $\beta$}})\; = $}}
\put(220,-5){\makebox(0,0){.}}

\end{picture} \\
Then the transfer matrix can be written as
\begin{equation}\label{e.t5v}
{\bf T}_{\mbox{{\tiny 5-v}}} = {\bf T}_{R} + {\bf T}_{L}\ \ \ .
\end{equation}

 From the ice rule, the number of up arrows on a row and right
arrows on a column are conserved. In the language of domain wall,
the number of up arrows per row corresponds to the number of domain
walls per row and the number of right arrows per column
corresponds to the number of type 1 domain walls per column. We will
call them $Q$ and $\Omega$, respectively.
 From the conservation of $Q$, ${\bf T}_{\mbox{{\tiny 5-v}}}$ is a
direct sum of submatrices labeled by $Q$ which only act on the
subspace with $Q$ domain walls. Thus,
\begin{equation}\label{e.t5vq}
{\bf T}_{\mbox{{\tiny 5-v}}} =
\bigoplus_{Q=0,\cdots,N} \left( {\bf T}_{R,Q} + {\bf
T}_{L,Q} \right)
\end{equation}
where $\bigoplus$ stands for the direct sum and ${\bf T}_{R,Q}\
({\bf T}_{L,Q})$ denotes the sector $Q$ of ${\bf T}_{R}\
({\bf T}_{L})$.

The partition function ${\cal Z}_{\mbox{{\tiny 5-v}}}$ of the
5--vertex model is obtained from the partition function
${\cal Z}_{\mbox{{\tiny TAFIM}}}$ of the $T=0$ TAFIM as follow.
Suppose the triangular lattice has $M$ rows and $N$
columns as in Fig.~2 under the boundary condition $(\mu,\nu)$
defined by
\begin{equation}\label{e.bcond}
\begin{array}{ccl}
s_{i,M+1}&=&(-1)^{\mu}s_{i,1}\hspace{15mm} (\mu =0,1)\\ [3mm]
s_{N+1,j}&=&(-1)^{\nu}s_{1,j}\hspace{15mm} (\nu =0,1)
\end{array}
\end{equation}
where $\mu$, $\nu$ are $0$ $(1)$ for periodic (anti-periodic)
boundary condition.
Let $\hbox{\boldmath{$s$}}=(s_1,\cdots,s_N)$ denote the spin
state of one row. It can also be represented
by $(s_1,{\hbox{\boldmath $\alpha$}})$ where
${\hbox{\boldmath $\alpha$}}=(\alpha_1,\cdots,\alpha_N)$ and
$\alpha_i = -s_is_{i+1}$. With the identification of $\alpha =
1\ (-1)$ to the up (down) arrow in the $i$-th vertical bond in the
dual lattice, one notes that ${\hbox{\boldmath $\alpha$}}$ is the
arrow configuration of a row of vertical bonds of the corresponding
5--vertex model configuration.
The transfer matrix ${\bf T}_{\mbox{{\tiny TAFIM}}}^{(\nu)}$ of the
$T=0$ TAFIM is defined through its matrix element
${\bf T}_{\mbox{{\tiny TAFIM}}}^{(\nu)}
(s_1,{\hbox{\boldmath $\alpha$}} \mid s_1',
{\hbox{\boldmath $\alpha$}}')$ which is the Boltzmann weight for
two successive row configurations
$(s_1,{\hbox{\boldmath $\alpha$}})$ and
$(s_1',{\hbox{\boldmath $\alpha$}}')$ with the boundary condition
$\nu$ along the horizontal direction. Due to the global spin
reversal symmetry, it takes the block form
\begin{equation}\label{e.ttisim}
{\bf T}_{\mbox{{\tiny TAFIM}}}^{(\nu)} = \left(
\begin{array}{cc}
{\bf T}_{\mbox{\tiny ++}} & {\bf T}_{\mbox{\tiny +{--}}} \\
{\bf T}_{\mbox{\tiny {--}+}} & {\bf T}_{\mbox{\tiny {--}{--}}}
\end{array}
\right) \sim \left(
\begin{array}{cc}
{\bf T}_{\mbox{\tiny ++}} + {\bf T}_{\mbox{\tiny +{--}}} & 0 \\
0 & {\bf T}_{\mbox{\tiny ++}} - {\bf T}_{\mbox{\tiny +{--}}}
\end{array} \right)
\end{equation}
where ${\bf T}_{s s'}\  (s,s'=\pm)$ is the matrix whose elements are
${\bf T}_{\mbox{{\tiny TAFIM}}}^{(\nu)}(s,{\hbox{\boldmath
$\alpha$}} \mid s',{\hbox{\boldmath $\alpha$}}')$ and $\sim$ denotes
an equivalence up to the similarity transformation. We use the fact
that $ {\bf T}_{s s'} = {\bf T}_{{-}s {-}s'} $.
If we denote the partition function of the $T=0$ TAFIM under the
boundary condition $(\mu,\nu)$ as
${\cal Z}_{\mbox{{\tiny TAFIM}}}^{(\mu,\nu)}$, it can be written as
\begin{equation}
{\cal Z}_{\mbox{{\tiny TAFIM}}}^{(\nu,\mu)} =
\mbox{Tr}\ {\bf R}^{\mu}\
\left[{\bf T}_{\mbox{{\tiny TAFIM}}}^{(\nu)}\right]^M
\end{equation}
where ${\bf R}$ is the spin reversal operator.

Since the sign of spin reverses by crossing each domain wall in
the horizontal direction, spin configurations under the boundary
condition $\nu=0\ (1)$ yield only domain wall
configurations with $Q$ even (odd).
Therefore, ${\bf T}_{\mbox{\tiny ++}}$
and ${\bf T}_{\mbox{\tiny +--}}$ in Eq.~(\ref{e.ttisim}) are
${\bf T}_{R}$ and ${\bf T}_{L}$, respectively, of the 5--vertex
model restricted to $Q$ even (odd) sectors for $\nu=0\ (1)$.
This shows that the transfer matrix spectra of the two models are
not identical.
Only the even or odd $Q$ sector of Eq.~(\ref{e.t5vq}) are
present in Eq.~(\ref{e.ttisim}) while the latter includes the block
${\bf T}_{\mbox{\tiny ++}}-{\bf T}_{\mbox{\tiny +--}}$ which is not
present in the 5--vertex model. And the spin configurations under
the boundary condition $\mu=0\ (1)$ yield only domain wall
configurations with $\Omega$ even (odd) assuming that $M$ is even,
since the sign of spin changes in every step except when crossing
a type 1 domain wall in the vertical direction.
If $M$ is odd, even (odd) $\Omega$ corresponds to $\mu=1$ $(0)$.
So, the partition function ${\cal Z}_{\mbox{{\tiny 5-v}}}$ of the
5--vertex model is given by
\begin{equation}\label{e.tafim.5v}
{\cal Z}_{\mbox{{\tiny 5-v}}} = \frac{1}{2} \left(
{\cal Z}_{\mbox{{\tiny TAFIM}}}^{(0,0)}+{\cal Z}_{
\mbox{{\tiny TAFIM}}}^{(0,1)}+{\cal Z}_{\mbox{{\tiny TAFIM}}}^{(0,1)}
+{\cal Z}_{\mbox{{\tiny TAFIM}}}^{(1,1)}
\right)
\end{equation}
where the factor $1/2$ accounts for
the two-to-one correspondence. This relation will be used in
Sec.~\ref{sec4} to obtain the toroidal partition function
$\widetilde{\cal Z}_{\mbox{{\tiny 5-v}}}$ of the
5--vertex model.

\section{Phase Diagram}\label{sec3}
The five--vertex model transfer matrix can be diagonalized by the
Bethe ansatz method as a special case of the general six--vertex
model~\cite{LieWu}.
Its phase diagram has recently been calculated by Gul\'{a}csi {\it
et al.}~\cite{GulBL} for the special case of $w_1=w_2$
\footnote{Gul\'{a}csi {\it et al.} use the notation $w_2=0$. Their
work and ours are related by the transformation $w_1
\leftrightarrow w_4, w_2 \leftrightarrow w_3$ and $ w_5
\leftrightarrow w_6 $.}.
In this section, we generalize it to the full three dimensional
parameter space and also calculate the domain wall densities. We
also discuss types of solutions of the Bethe ansatz equation (BAE)
of the 5--vertex model.

The eigenvalues of the transfer matrix Eq.~(\ref{TM}) in the sector
$Q\ (\neq N)$ are given by~\cite{LieWu,GulBL}
\begin{equation}\label{e.evdw}
\Lambda_Q = w_2^{N-Q} w_4^Q \prod_{j=1}^Q \left( 1 +
\frac{w_5 w_6}{w_2w_4} z_j \right)
\end{equation}
where the set $\{z_1,z_2,\ldots,z_Q\}$ are the solutions of the BAE
\begin{equation}\label{e.baez}
z_j^N = (-1)^{Q-1} \prod_{l=1}^Q \frac{1-\Delta z_j}{1-\Delta z_l}
\ \ ,\ \ j=1,2,\ldots,Q
\end{equation}
with
\begin{equation}\label{e.delta}
\Delta = \frac{w_1w_2-w_5w_6}{w_2w_4} \ \ \ .
\end{equation}
All $z_j$'s should be distinct. When $Q=N$, $\Lambda_N=w_1^N+w_4^N$.

An alternative expression for the eigenvalue which is useful for
$Q>N/2$ is given by
\begin{equation}\label{e.evdh}
\Lambda_Q=w_1^Q\prod_{j=1}^{\bar{Q}}\left\{\frac{w_5w_6}{w_1z_j-w_4}
\right\} + w_4^Q \prod_{j=1}^{\bar{Q}}
\left\{w_2-\frac{w_5w_6z_j}{w_1z_j-w_4}\right\}
\end{equation}
where $\bar{Q}\equiv N-Q$ is the number of domain wall holes and
the set $\{z_1,z_2,\ldots,z_{\bar{Q}}\}$ is again given by
Eq.~(\ref{e.baez}) with $Q$ replaced by $\bar{Q}$.
We call Eq.~(\ref{e.evdw}) and Eq.~(\ref{e.evdh}) the domain wall
representation and the domain wall hole representation, respectively.
Using Eq.~(\ref{e.vw}) into Eqs.~(\ref{e.evdw}), (\ref{e.delta}) and
(\ref{e.evdh}) gives $\Lambda_Q$ in terms of the domain wall
parameters as
\addtocounter{equation}{-4}
\renewcommand{\theequation}{\arabic{equation}'}
\begin{equation}\label{e.evdwp}
\Lambda_Q= x_2^Q \prod_{j=1}^Q\left( 1 + a z_j\right)
\end{equation}
or
\addtocounter{equation}{2}
\begin{equation}\label{e.evdhp}
\Lambda_Q = x_2^Q (\Delta+a)^Q \prod_{j=1}^{\bar{Q}}\left\{
\frac{a}{(\Delta+a)z_j-1}\right\} + x_2^Q \prod_{j=1}^{\bar{Q}}
\left\{1-\frac{az_j}{(\Delta+a)z_j-1}\right\}
\end{equation}
where  $a$ is the ratio of two domain wall fugacities
\renewcommand{\theequation}{\arabic{equation}}
\begin{equation}
a = x_1 / x_2
\end{equation}
and
\addtocounter{equation}{-3}
\renewcommand{\theequation}{\arabic{equation}'}
\begin{equation}
\Delta = x_1(y_1-1)/x_2 = a(y_1-1)\ \ \ .
\end{equation}
\addtocounter{equation}{2}
\renewcommand{\theequation}{\arabic{equation}}

\noindent\noindent
We will call $\Delta$ the interaction parameter. It is positive for
attractive interaction and negative for repulsive interaction
between domain walls.

Defining the momenta $\{p_j\}$ by $z_j  = e^{i p_j}$,
Eq.~(\ref{e.baez}) also takes the familiar form
\begin{equation}\label{e.baep}
N p_j = 2 \pi I_j + \sum_{l=1}^Q \Theta(p_j,p_l)
\end{equation}
where
\begin{equation}\label{e.theta}
e^{i\Theta(p,q)} \equiv
-\displaystyle{\frac{1-\Delta e^{i p}}{1-\Delta e^{i q}}}
\end{equation}
and $I_j$'s are half-integers for even $Q$ and integers for odd $Q$
ranging from $-(N-1)/2$ to $(N-1)/2$. Different eigenvalues come
from different choices of the set $\{I_j\}$.

The BAE may take another form. If we define
\begin{equation}\label{e.sdef}
s=\frac{1}{Q} \sum_l \ln(1-\Delta z_l)\ \ \ ,
\end{equation}
then the BAE becomes
\begin{equation}
z_j = (-1)^{(Q-1)/N} \left( 1 - \Delta z_j \right)^q e^{-qs}
\end{equation}
where $q=Q/N$ is the domain wall density.
This equation gives $z_j$'s as a function of $s$ which should,
in turn, be determined from its defining equation~(\ref{e.sdef}).

Note that the BAE (Eq.~(\ref{e.baez})) arises from the periodic
boundary condition on the wave function of
${\bf T}_{\mbox{{\tiny 5-v}}}$~\cite{IzySkr}.
It is also interesting to consider another boundary condition,
say, the anti-periodic boundary condition. The effect of the
boundary condition is to shift domain walls out of the $N$-th
site to the first site with appropriate phase factor $1\ (-1)$
for periodic (anti-periodic) boundary condition.
The shift operation is done by the operator ${\bf T}_R$. So, if
we impose anti-periodic boundary condition, the resulting matrix we
diagonalize is ${\bf T}_{L}-{\bf T}_{R}$. In this case, the
expression for eigenvalues remains the same except the fact that
$I_j$ should be integers for even $Q$ and half-integers for odd $Q$.
So, we can obtain whole spectrum of the transfer matrix of the
$T=0$ TAFIM from the transfer matrix of the 5--vertex model under
periodic and anti-periodic boundary conditions. Note that the
anti-periodic boundary condition here is different from that which
reverses the sense of horizontal arrows.

The free energy in the language of the domain wall physics is a
function of $x_1,x_2$ and $y_1$ through Eq.~(\ref{e.vw}). From now
on, we regard it as a function of $\Delta, x_2$ and $a=x_1/x_2$.
Since the free energy is given by the maximum eigenvalue of the
transfer matrix, $f(x_2,a,\Delta)$, the free energy per site in the
units of $kT$, is written in the form
\begin{eqnarray}
f(x_2,a,\Delta)&=&-\lim_{N\rightarrow\infty} \max_{Q} \left[
\frac{1}{N}\ln{\Lambda_Q}\right]\nonumber \\[3mm]
&=&- \max_{q} \left[q \ln{x_2}+\kappa(q) \right]
\end{eqnarray}
where $\kappa(q)$, which will be called the configurational free
energy, is given by
\begin{equation}\label{e.kpdef}
\kappa(q) = \lim_{N\rightarrow\infty} \max_{\{z_j\}}
\left[ \frac{1}{N}\sum_{j=1}^Q\ln{\left(1+ a  z_j\right)} \right]\ .
\end{equation}
Here, $\{z_j\}$'s are the solutions of the {\em BAE}.
The equation of state which relates the equilibrium domain wall
density $q$ as a function of thermodynamic parameters is given by
the relation
\begin{equation}
q(x_2,a,\Delta) = \frac{Q_0}{N}
\end{equation}
where $Q_0$ is the value of $Q$ at which $\Lambda_Q$ attains  the
maximum value. The equation of state can be rewritten as
\begin{equation}\label{e.eos}
\frac{\partial}{\partial q}\kappa(q,a,\Delta) = - \ln{x_2}
\end{equation}
if $\kappa(q)$ is a differentiable and convex function.
The configurational free energy $\kappa(q)$ is a Legendre
transformation of $f$. That is, it is a free energy as a function
of domain wall density while $f$ is a free energy as a function of
the domain wall fugacity.

We now classify types of solutions of the BAE corresponding to the
maximum eigenvalue. First, consider the case $-1<\Delta<1$. This
region contains the non-interacting case with $\Delta=0$ which is
considered in \cite{LieWu}. In this case, the $\Theta$ function
defined in Eq.~(\ref{e.theta}) is identically 0. So, any set
$\{I_j\}$ of $Q$ different numbers are solutions of the BAE and the
solution giving the maximum value of $\Lambda_Q$ is $\{I_j\} =
\left\{-(Q-1)/{2},-(Q-1)/{2}+1,\ldots,(Q-1)/{2}\right\}$. We
assume that this set $\{I_j\}$ still gives the maximum eigenvalue
even after turning-on of weak interaction and remains so in the
whole region $-1<\Delta<1$. This assumption is tested by direct
numerical diagonalization of the transfer matrix with $N$ up to
15. We call this type of solution as the {\em free magnon type}.

When $|\Delta| > 1$, there appear other types of solutions. Assume
that the solution is of the form,
\begin{equation}
\left\{
\begin{array}{ccll}
z_j&=&\Delta^{b_j} \bar{z}_j & , \ \ \ j=1,\ldots,N_+ \\[3mm]
z_j&=&\bar{z}_j &,\ \ \ j=(N_+ {+} 1),\ldots, (N_+{+}N_0)\\[3mm]
z_j&=&\displaystyle{{\frac{1}{\Delta}}-{\frac{\bar{z}_j}
{\Delta^{1+a_j}}}} &, \ \ \ j=(N_+{+}N_0+1) ,\ldots, Q
\end{array} \right.
\end{equation}
where $a_j$ and $b_j$ are constants greater than 0 and $\bar{z}_j$'s
are assumed to remain of order 1 as $|\Delta| \rightarrow \infty$.
In other words, of $Q$ $z_j$'s, $N_+$ are diverging, $N_0$ remain
finite and $N_-=Q{-}(N_+{+}N_0)$ vanish inversely as $|\Delta|
\rightarrow \infty$. Then, the necessary condition that this set
should be a solution of the BAE is either (i) $N_+ = N_- = 0$ so
that all $z_j$'s are of order 1 or (ii) $N_0 = 0,b_j=(Q-N_+)/N_+
\mbox{ and } a_j = (N-Q)/N_+$. We also call the first type as the
{\em free magnon type} while the second type will be called as the
{\em bounded magnon type}. Of these possibilities, one can easily
show that the configurational free energy is realized by the {\em
free magnon type} if $\Delta \rightarrow -\infty$ and the {\em
bounded magnon type} with $N_+=1$ if $\Delta \rightarrow
\infty$. We find numerically this feature also persists for all
$\Delta$ in the range $|\Delta|>1$.

It is very difficult to obtain the equation of
state~(Eq.~(\ref{e.eos})) analytically for whole range of
parameters $x_2, a$ and $\Delta$. But, the phase boundaries which
separate the commensurate phases with domain wall
density $0$ or $1$ from the incommensurate phase can be obtained
if we solve the BAE in the $q \rightarrow 0$ or 1 limit.
Apart from the C phases with domain wall density 0 or 1, there
appears new a C phase with $q=1/2$ if $a$ is large so that $\Delta$
can take values less than some critical value $\Delta_c$. (Below, we
will see that $\Delta_c$ takes the value $-4$.)
 Consider the case where domain
walls of type 1 are much more favorable to form than those of type
2 ($a=x_1/x_2 \gg 1$) and there are repulsive
interaction between them ($y_1<1$) so that the interaction parameter
$\Delta$ is less than $\Delta_c$.
Then, the most energetically favorable
state for $q\leq 1/2$ is the state where there are only type 1
domain walls with no adjacent pairs. But, if $q$ is larger
than $1/2$, there should appear type 2 walls and adjacent pairs of
domain walls of type 1 whose energy costs are large.
So, it is expected that there is a discontinuity in $x_2$ which
controls the total number of domain
walls across the $q=1/2$ line.
The phase boundary of the $q=1/2$ C phase can be also determined
analytically. Our results for the phase boundary are given by
Eqs.~(\ref{e.x0c}), (\ref{e.x1cl}), (\ref{e.x1cs}), (\ref{e.xc}),
(\ref{z.d}) and (\ref{z.u}) and are illustrated in Fig.~7.

\noindent
\underline{(i) $\Delta < 1,\ q=0$}

In the region $\Delta <1$, the {\em free magnon type} solution in
the $q \rightarrow 0$
limit is \begin{equation}\label{e.solq0}
z_j=e^{i \theta_j} \left[1+\frac{\Delta q}{1-\Delta} \left(
1-e^{i\theta_j}\right)\right] + {\cal O}(q^3)
\end{equation}
where $\theta_j=2\pi I_j/N$ and $\{I_j\}$ is a set of integers or
half-integers depending on
the parity of the domain wall number $Q$. The maximum value of
$\kappa(q)$ is obtained if we choose the set $\{I_j\}=\{-(Q-1)/2,
-(Q-1)/2+1,\ldots,(Q-1)/2\}$ and the next largest values of
$\kappa(q)$ are obtained by using the set $\{I_j'=I_j+m\}$ which
is a shift of the set $\{I_j\}$ by an integer $m$.
With this solution, the configurational free energy $\kappa$ is
given from Eq.~(\ref{e.kpdef}) by
\begin{eqnarray}\label{e.kp0dl}
\kappa(q)
&=& \frac{1}{2\pi} \int_{-\pi q}^{\pi q} \ln\left[ 1 + a e^{i\theta}
\left( 1 + \frac{\Delta q}{1-x} ( 1 - e^{i\theta})\right) \right]
d\theta + {\cal O}(q^4) \nonumber \\
&=& q\ln(1+a) - \frac{a\pi^2}{6(1+a)} q^3 + {\cal O}(q^4) \ \ \ .
\end{eqnarray}
This, together with Eq.~(\ref{e.eos}), implies that $q=0$ if $x_2
\leq 1/(1+a)$. Thus, we obtain the phase boundary $x_2=x_{0C}$
between the $q=0$ C phase and the IC phase as
\begin{equation}\label{e.x0c}
x_{0C} = \frac{1}{1+a}
\end{equation}
or equivalently $x_1+x_2 =1$. For $x_2$ slightly larger that
$x_{0C}$, Eq.~(\ref{e.kp0dl}) gives
\begin{equation}
q \simeq \sqrt{\frac{2(1+a)}{a\pi^2}}\ \left( \ln{x_2} -
\ln{x_{0C}} \right)^{1/2}
\end{equation}
The domain wall density thus shows the square root dependence on
domain wall formation energy which is the general character of
the PT transition. This type of singularity is originated from the
fact that the leading contribution $\kappa(q)$ aside from the linear
term is of order $q^3$.
It is originated from the entropy reduction due to the collision of
domain walls~\cite{Fis}.

\noindent
\underline{(ii) $\Delta<1,\ q=1$}

Next, consider the case near $q=1$. In this case, it is easier to
consider the BAE for domain wall hole rather than domain wall.
Inserting Eq.~(\ref{e.solq0}) with $q$ replaced by $\bar{q}$
into Eq.~(\ref{e.evdhp}), we obtain $\kappa(q)$ near $q=1$.
$\bar{q}=1-q$ is a domain wall hole density.
There are two cases to consider depending on whether $\Delta+a>1$ or
$\Delta+a<1$. When $\Delta+a>1$, the first term in the right hand
side of Eq.~(\ref{e.evdhp}) dominates and hence
\begin{eqnarray}
\kappa(q)&=&q\ln(\Delta+a)+\frac{1}{N} \sum_{j=1}^{\bar{Q}}\ \ln
\left[\frac{a}{(\Delta+a)z_j-1}\right]\nonumber \\
d\theta&=&\ln(\Delta+a)-\bar{q}\ \ln\frac{(\Delta+a)(\Delta+a-1)}{a} -
\frac{\bar{q}^3}{6} \frac{\pi^2 (\Delta+a)}{(\Delta+a-1)^2} +
{\cal O}(\bar{q}^4)\ \ .
\end{eqnarray}
 From this, the phase boundary $x_2=x_{1C}$ between $q=1$ C phase and
IC phase is given by
\begin{equation}\label{e.x1cl}
x_{1C} = \frac{a}{(\Delta+a)(\Delta+a-1)} \hspace{15mm}(\Delta+a>1)
\end{equation}
and the equilibrium domain wall density near $x_{1C}$ is given by
\begin{equation}
q \simeq 1 - \sqrt{\frac{2(\Delta+a-1)^2}{\pi^2(\Delta+a)}} \left(
\ln{x_{1C}}-\ln{x_2} \right)^{1/2} \ \ \ .
\end{equation}
Similarly, when $\Delta+a<1$, the configurational free energy is
given by
\begin{eqnarray}
\kappa(q)&=&\frac{1}{N}\sum_{j=1}^{\bar{Q}}\ln\left[1-\frac{a}{
(\Delta+a)z_j-1} \right] \nonumber \\[3mm]
&=&\bar{q}\ln\frac{1-\Delta}{1-\Delta-a}-\frac{1}{6}\bar{q}^3
\frac{a\pi^2(1-a\Delta-\Delta^2)}{(\Delta+a-1)^2 (\Delta-1)^2}
+{\cal O}(\bar{q}^4) \ \ \ ,
\end{eqnarray}
from which the phase boundary $x_2=x_{1c}$ and the equilibrium
domain wall density near $x_{1C}$ are given by
\begin{equation}\label{e.x1cs}
x_{1C} = \frac{1-\Delta}{1-\Delta-a} \hspace{15mm} (\Delta+a < 1)
\end{equation}
and
\begin{equation}
q \simeq 1 - \sqrt{\frac{2(\Delta+a-1)^2(\Delta-1)^2}
{a\pi^2(1-a\Delta- \Delta^2)}} \left( \ln{x_{1c}} - \ln{x_2}
\right)^{1/2} \ \ \ .
\end{equation}
So, we conclude that when $\Delta <1$, there are commensurate
phases with domain wall density 0 for $x_2 < x_{0c}$ and domain
wall density 1 for $x_2 > x_{1c}$. In between, the equilibrium
domain wall density increase smoothly as $x_2$ increases as long
as $\Delta>-4$. Fig.~6 shows a typical $x_2$ dependence of $q$
for the case of $a=1$.
The curves are obtained numerically by solving the BAE for $N$ up to
150. The case $\Delta<-4$ will be considered later.

\noindent
\underline{(iii)\ $\Delta > 1 $}

Now, consider the case where there is a strong attractive
interaction between domain walls so that $\Delta >1$.
The solution of the BAE maximizing $\kappa$ is of the {\em
bounded magnon type} with $N_+ =1$. The exact solution of the BAE is
easily obtained from the transformation of $z$ to $\bar{z}$ defined
by
$$ \left[
\begin{array}{ccl}
z_1 &=& \bar{z}_1 \Delta^{(Q-1)} \\
z_{j\neq 1} &=& \displaystyle{
\frac{1}{\Delta}} \left(1-\displaystyle{ \frac{\bar{z}_j}{\Delta^{N-Q}}}
\right) \ \ \ .
\end{array}
\right. $$
Then the BAE for $\bar{z}_j$ becomes
\begin{equation}
\left[
\begin{array}{ccl}
\bar{z}_1^N &=& \displaystyle{
\frac{\left( \bar{z}_1 - \Delta^{-Q}\right)^{Q-1}}{
\prod_{l\neq 1}^Q \bar{z}_j }} \\ [3mm]
\left(1- \displaystyle{
\frac{\bar{z}_j}{\Delta^{N-Q}}} \right)^N&=&\displaystyle{(-1)^{Q-1}
\frac{\bar{z}_j^{Q-1}}{\left(\Delta^{-Q}-\bar{z}_1\right)
\prod_{l\neq 1,j} \bar{z}_l }} \ \ \ .
\end{array}
\right.
\end{equation}
For macroscopic number of $N$ and $Q$, the values of
$\bar{z}_j/\Delta^{N-Q}$ and
$\Delta^{-Q}$ are exponentially small and may be neglected.
Thus the solution is
\begin{equation}\label{e.sol}
\left[ \begin{array}{ccl}
\bar{z}_1 &=& e^{i\theta} \\[3mm]
\bar{z}_{j\neq 1} &=& \bar{z}_1 e^{i\pi (1+2j/Q)}
\end{array}
\right.
\end{equation}
where $\theta$ can take the value from the set $\frac{2\pi}{N}
\times \{1,\ldots,N\}$.
The corresponding right-eigenvector $|\lambda\rangle$ of the
${\bf T}_{R,Q}+{\bf T}_{L,Q}$ is
\begin{equation}\label{e.kbdw}
|\lambda\rangle = C \sum_{n_1<\cdots<n_Q}e^{in_1\theta}
\Delta^{-\sum_{l=2}^Q (n_l-n_1)} |n_1,\ldots,n_Q\rangle
\end{equation}
where $C$ is a normalizing constant. Here, $\{n_i\}$'s denote the
position of up arrows or equivalently domain walls.
It is obvious that these states represent bounded domain wall states
because the components of
eigenvector decay exponentially as the distance between domain walls
becomes large. In fact, one can calculate the mean distance of the
last domain wall from the first one.
The mean distance $\langle n_Q-n_1 \rangle$ of $Q$ domain wall
system is given by
\begin{equation}\label{e.sbdw}
\langle n_Q-n_1\rangle=\frac{\langle \lambda |\left( \hat{n}_Q -
\hat{n}_1\right)|\lambda\rangle}{\langle\lambda |\lambda \rangle}
\end{equation}
where $\hat{n}_j$ is the position operator of the $j$-th domain
wall. Inserting the eigenket to the above expression and after some
algebra, we find that $\langle n_{Q}-n_{1} \rangle$ is equal to
$Q$ for macroscopic number of domain walls.  Thus, we can
interpret this state as the bounded domain wall state.

This solution yields the exact configurational free energy which
is obtained from the solution Eq.~(\ref{e.sol}) with $\theta=0$;
\begin{equation}
\kappa(q) = q \ln(\Delta+ a ) \ \ \ .
\end{equation}
And the free energy $f$ for $\Delta > 1$ is simply
\begin{equation}
-\beta f=\max_{0\leq q\leq 1}
\left[ q\ln{x_2}+q\ln(\Delta+ a )\right] \ \ \ .
\end{equation}
The maximum value is obtained at $q=0$ if $x_2$ is less than
$1/(\Delta+ a )$ and at $q=1$ if $x_2$ is greater than
$1/(\Delta+ a )$. So there is a first order phase transition between
the two commensurate phases
when $x_2$ is at the critical fugacity $x_c$, where
\begin{equation}\label{e.xc}
x_c = \frac{1}{\Delta+ a } \ \ \ \ \ \ ( \Delta > 1 ) \ \ \ .
\end{equation}
Note that the condition $w_1=w_2$ used in \cite{GulBL} is amount to
the condition $x_2=1/(\Delta+a)$ so that the first order transition
for $\Delta>1$ could not to be seen in \cite{GulBL}.
We have thus found the phase boundary of the C phase with domain
wall density $q=0$ and 1 and the nature of the phase transition. We
present the resulting phase diagram  in Fig.~7(a) for the case of
$a=1$.

%%%%%%%%%%%%%%%%%%%%%%%%%%%%%%%%%%%%%%%%%%%%%%%%%%%%%%%%%%%%%%%%%%%%
\noindent\noindent
\underline{(iv) $\Delta < -4$, $ q = 1/2^{-}$}

As discussed before, we expect that $f$ has a singularity in $x_2$ at
$q=1/2$ if $\Delta$ is large and negative.
To see $q$ dependence of $x_2$ near $q=1/2$, we should evaluate
the configurational free energy $\kappa(q)$ near $q=1/2$.
Gul\'{a}csi {\it et al.}~\cite{GulBL} used the root density
function $\rho(p)$ to find the $q=1/2$ phase boundary when
$x_2(\Delta+a)=1$. We employ the same method to the general case.

$\rho(p)$ is defined so as
$N\rho(p) dp$ to be the number of the roots of the BAE
(Eq.~(\ref{e.baez})) with $z=e^{ip}$ in the interval $(p,p+dp)$ in
the complex $p$ plane. We stress here that the roots do not lie on a
straight line in the complex $p$ plane. For
domain wall density $q$, $\rho(p)$ is given by~\cite{GulBL}
\begin{equation}
\rho(p) = \frac{1}{2\pi} \left( 1 + q \frac{\Delta e^{ip}}{
1-\Delta e^{ip}} \right) \ \ \ .
\end{equation}
In Fig.~8, we give a typical root distribution of the BAE in the
complex $\alpha$ plane which is related to $p$ as
\begin{equation}\label{e.ptoal}
e^{i\alpha} \equiv 1 - \Delta e^{ip}\ \ \ .
\end{equation}
The root density function $\tilde{\rho}(\alpha)$ in the $\alpha$
plane is given by
\begin{equation}
\tilde{\rho}(\alpha) \equiv  \rho(p) \frac{dp}{d\alpha} =
\frac{1}{2\pi} \left( -q + \frac{e^{i\alpha}}{e^{i\alpha}-1} \right)
\end{equation}
and the variables $A$, $B$ and $D$ in Fig.~8 are given by
\begin{equation}
\left\{
\begin{array}{l}
q =\displaystyle{ \mbox{ Im}\left[ \ln{(e^{iA-B}-1)}
\right]/(\pi +A)}\\
\displaystyle{\mbox{Re}\left[\ln\frac{1-e^{iA-B}}{\Delta}\right] +
\frac{A}{\pi} \mbox{Re}
\left[ \ln{(e^{iA-B}-1)} \right] +\frac{AB}{\pi} + \frac{1}{\pi}
\sum_{n=1}^\infty \frac{e^{nB}}{n^2} \sin{nA} = 0} \\
D = 1 - \Delta C
\end{array}
\right.
\end{equation}
where $C$ is determined from the equation
\begin{equation}
C = \frac{(1-\Delta C)^q}{ \exp\left[ \frac{1}{N} \sum_{j=1}^Q
\ln(1- \Delta e^{ip_j}) \right] } \ \ \ .
\end{equation}
Near  $q=1/2$, they take the values
\begin{equation}
\left\{
\begin{array}{ccl}
A &=&\displaystyle{ \pi + (q-1/2) \frac{1}{4\pi}
\frac{1-\lambda}{1+\lambda}+{\cal O}((q-1/2)^2)} \\
B &=&\displaystyle{ \ln{\lambda} +{\cal O}((q-1/2)^2)} \\
D &=&\displaystyle{ 1 + \frac{|\Delta|}{2} \left( 1 +
\sqrt{1+\frac{4}{|\Delta|}} \right)
+(q-1/2)\ln{(d\lambda)}/\left(\frac{1}{d-1}-\frac{1}{2d}\right)
+{\cal O}((q-1/2)^2)}
\end{array}
\right.
\end{equation}
with
\begin{equation}
\lambda = \frac{1}{2} \left( |\Delta| - 2 - \sqrt{
(|\Delta|-2)^2 - 4 } \right)\ \ \ .
\end{equation}

With this knowledge, we can calculate
$\kappa(q)$ near $q=1/2$. First consider the case $q<1/2$, where
$\kappa(q)$ is evaluated from Eq.~(\ref{e.evdwp}).
\begin{eqnarray}
\kappa(q) &=& \frac{1}{N} \sum_{j=1}^Q \ln\left( 1 + a e^{ip_j}
\right) \nonumber \\
&=& \int_{\cal C} d\alpha\ \tilde{\rho}(\alpha)\
\ln \left( \frac{ a+\Delta-ae^{i\alpha}}{\Delta} \right)
\end{eqnarray}
where the integration should be taken along the contour ${\cal C}$
shown in Fig.~8. But, the contour can be deformed to the straight
line ${\cal L}$ since the integrand is analytic in the shaded
region. Then $\kappa(q)$, up to the first order in $(q-1/2$), is
given by
\begin{eqnarray}
\kappa(q) &=& \frac{1}{2\pi} \int_{-A}^A \left[ -q + \frac{ e^{it}/
\lambda}{e^{it}/\lambda-1}\right] \ln\left( \frac{a+\Delta-ae^{it}/
\lambda}{ \Delta} \right) \nonumber \\
&=& \frac{1}{2}\ln{a} + (q-1/2) \left[ \ln\frac{a}{|\Delta|\lambda}
+ 2 \ln\left(1+\frac{a+\Delta}{a}\lambda\right) \right] + {\cal O}
((q-1/2)^2)\ \ \ .
\end{eqnarray}
Note that $\kappa(1/2) = \frac{1}{2}\ln{a}$ independent of $\Delta$.
This implies that there are
only type 1 domain walls without adjacent pairs of them at $q=1/2$
phase.  From the Eq.~(\ref{e.eos}), we see that $q=1/2$ phase starts
at $x_2= x_{-}$ where
\begin{equation}\label{z.d}
\ln{x_{-}} = -\ln\frac{a}{|\Delta|\lambda} - 2 \ln\left[ 1 +
\frac{a+\Delta}{a}\lambda \right] \ \ \ .
\end{equation}

\noindent\noindent
\underline{(v) $\Delta<-4$, $q=1/2^{+}$}

Next consider the case $q>1/2$, where the configurational free
energy is evaluated from the Eq.~(\ref{e.evdhp}). For convenience,
we define two quantities ${\cal A}$ and ${\cal B}$;
\begin{eqnarray}
{\cal A}&\equiv&\frac{1}{N}\sum^{\bar{Q}} \ln\left(1-\Delta
e^{ip_j}\right)\nonumber \\
{\cal B}&\equiv&\frac{1}{N}\sum^{\bar{Q}}
\ln\left(1-(a+\Delta)e^{ip_j}\right).
\end{eqnarray}
Then the configurational free energy is written as
\begin{equation}
\kappa(q) = \max\{\kappa_L(q),\ \kappa_R(q)\}
\end{equation}
where $\kappa_L= {\cal A} - {\cal B}$ and $\kappa_R=\ln{a} + q
\ln((a+\Delta)/a)-{\cal B}$.
The quantity ${\cal A}$ and ${\cal B}$ can be written as a contour
integration in the complex $\alpha$ plane in the thermodynamic
limit $N\rightarrow \infty$.
\begin{eqnarray}
{\cal A} &=& \int_{\cal C}d\alpha\ i\alpha\ \tilde{\rho}(\alpha)
\nonumber \\
{\cal B} &=& \int_{\cal C}d\alpha\ \tilde{\rho}(\alpha)\ \ln\left(
\frac{a-(a+\Delta)e^{i\alpha}}{|\Delta|}\right)  .
\end{eqnarray}
Since the integrand in ${\cal A}$ is analytic in the shaded region,
the contour can be deformed to the straight line  ${\cal L}$ and
the integration results in
\begin{equation}
{\cal A}=\frac{1}{2}\ln{|\Delta|}-(\bar{q}-\frac{1}{2}) \ln\lambda+
{\cal O} ((\bar{q}-1/2)^2)\ \ \ .
\end{equation}
In other to calculate ${\cal B}$, there are three possible cases to
consider;

\noindent\noindent
(a) $1/D <\ \lambda <\ (a+\Delta)/a$

The integrand is also analytic in the shaded region and the contour
${\cal C}$ is replaced by the straight line ${\cal L}$. It yields
\begin{equation}
{\cal B}_a = \frac{1}{2} \ln(a+\Delta) + (\bar{q}-1/2)\left[
\ln\frac{a+\Delta}{|\Delta|\lambda}+2\ln\left( 1 + \frac{a\lambda}{
a+\Delta}\right)\right]+{\cal O}\left((\bar{q}-1/2)^2 \right)\ \ \ .
\end{equation}

\noindent\noindent
(b) $1/D <\ (a+\Delta)/a <\ \lambda$

In this case, the branch cut intrudes into the shaded area.
Therefore, upon
changing ${\cal C}$ to the straight line, one need to subtract the
contribution around the branch cut. The results is
\begin{equation}
{\cal B}_b = \frac{1}{2} \ln{(a+\Delta)} + (\bar{q}-1/2) \left[
\ln\frac{a+\Delta}{|\Delta|\lambda} +  2
\ln\left( 1 + \frac{a\lambda}{ a+\Delta}\right) \right]+
{\cal O}\left( (\bar{q}-1/2)^2 \right) \ \ \ .
\end{equation}

\noindent\noindent
(c) $(a+\Delta)/a <\ 1/D < \lambda$

In this case, the contour can be deformed to the straight line
${\cal L}$ as in the case (a) and the integration results in
\begin{equation}
{\cal B}_c = \frac{1}{2} \ln\frac{|\Delta|}{a} + (\bar{q}-1/2)
\left[ \ln\frac{a}{|\Delta|} + 2 \ln\left( 1 + \frac{a+\Delta}
{a\lambda}\right)\right]+{\cal O}\left((\bar{q}-1/2)^2 \right)\ \ \ .
\end{equation}

For each case, the quantity $\kappa_R$ and $\kappa_L$ take the
following values at $q=1/2$;
\begin{equation}\label{e.abc}
\left\{
\begin{array}{ccclcccl}
{\rm (a)}&\kappa_L &=& \frac{1}{2} \ln\frac{|\Delta|}{a+\Delta} &,&
\kappa_R&=&\frac{1}{2} \ln{a} \\ [3mm]
{\rm (b)}&\kappa_L &=& \frac{1}{2} \ln\frac{|\Delta|}{a+\Delta} &,&
\kappa_R&=&\frac{1}{2} \ln{a} \\ [3mm]
{\rm (c)} & \kappa_L &=& \frac{1}{2} \ln{a} &,&
\kappa_R &=& \frac{1}{2} \ln{a} + \frac{1}{2}
\ln\frac{a(a+\Delta)}{|\Delta|} \ \ \ .
\end{array} \right.
\end{equation}

When $a+\Delta>1$, only the cases (a) and (b) in Eq.~(\ref{e.abc})
occur and $\kappa_R > \kappa_L$ at $q=1/2$. So near $q=1/2$, the
configurational free energy $\kappa(q)$ is
determined by $\kappa_R(q)$ and is given by
\begin{equation}\label{e.kpg}
\kappa(q)
= \frac{1}{2} \ln{a} + (q-1/2) \left[ \ln\frac{\lambda a}{|\Delta|}
+ 2\ln\left( 1+\frac{a+\Delta}{a\lambda} \right) \right] +{\cal
O}((q-1/2)^2)
\end{equation}

When $a+\Delta<1$, the cases (b) and (c) occur. One can easily find
that if $(a+\Delta)/a < 1/D$ then  $\kappa_L > \kappa_R$ and
if $(a+\Delta)/a > 1/D$ then $\kappa_L < \kappa_R$.
So, near $q=1/2$ the configurational free energy is
\begin{eqnarray}\label{e.kpl}
\kappa(q) &=& \left\{
\begin{array}{ll}
\kappa_L(q) \ \ \ ,&\ \mbox{if}\ \ (a+\Delta)/a < 1/D \\ [3mm]
\kappa_R(q) \ \ \ ,&\ \mbox{if}\ \ (a+\Delta)/a > 1/D
\end{array} \right. \nonumber \\
&=& \frac{1}{2} \ln{a} + (q-1/2) \left[ \ln\frac{\lambda a}{|\Delta|}
+ 2\ln\left( 1+\frac{a+\Delta}{a\lambda} \right) \right] +{\cal
O}((q-1/2)^2)
\end{eqnarray}
 From the Eq.~(\ref{e.kpg}) and (\ref{e.kpl}), we see that the
$q=1/2$ phase ends at $x_2=x_{+}$ where
\begin{equation}\label{z.u}
\ln{x_{+}} = -\ln\frac{a\lambda}{|\Delta|} - 2\ln\left( 1 +
\frac{a+\Delta}{a\lambda}\right)
\end{equation}

$x_{-}$ (Eq.~(\ref{z.d})) and $x_{+}$ (Eq.~(\ref{z.u})) defines the
phase boundary between the $q=1/2$ C phase and the IC phase and the
domain wall density is locked at $q=1/2$ for the range
\begin{equation}
x_{-} < x_2 < x_{+} \ \ \ .
\end{equation}
Since $x_{-}$ and $x_{+}$ merge at $\Delta_c=-4$, this phase appears
only when $a>4$.  Fig.~7(b) shows the full phase boundaries for
$a=7$ and Fig.~9 shows the domain wall density as a function of
$\ln{x_2}$ for $\Delta=-4$ and $-5$.

\section{The Critical Properties of the IC Phase}\label{sec4}
The conformal field theory predicts that the operator content of a
critical phase is related to the finite size correction to the
eigenvalue spectra of the transfer matrix~\cite{Carb}. When we
write an eigenvalue $\Lambda_\alpha$ of the transfer matrix for a
lattice of width $N$ as $e^{-E_\alpha}$, then $E_{\alpha}$ takes
the form at the criticality,
\begin{equation}\label{e.fss}
E_\alpha = N f_\infty + \frac{2\pi}{N} \left( \Delta_{\alpha} +
\bar{\Delta}_{\alpha} - \frac{c}{12} \right) \zeta \sin{\theta} +
\frac{2\pi i}{N}\left(\Delta_{\alpha}-\bar{\Delta}_{\alpha} \right)
\zeta \cos{\theta} + o\left( \frac{1}{N} \right) \ \ \ ,
\end{equation}
where $c$ is the central charge, $\left(\Delta_{\alpha},
\bar{\Delta}_{\alpha}\right)$ are the conformal dimensions of the
operator corresponding to the $\alpha$-th energy eigenstate,
$\zeta$ is the anisotropy factor, $\theta$ is the anisotropy angle
and finally $f_\infty$ is the non-universal bulk free energy per
site in units of $kT$~\cite{Carb,KimP}.

The {\em toroidal partition function} (TPF) $\widetilde{\cal Z}$ is
defined as the order 1 part of the partition function ${\cal Z}$ for
conformally invariant system of $N$ columns and $M$ rows. It follows
from Eq.~(\ref{e.fss}) that
\begin{eqnarray}
\widetilde{\cal Z} &\equiv& \lim_{\stackrel{N,M\rightarrow\infty}
{M/N = fixed}}
\sum_{\alpha} e^{-M E_\alpha}/ e^{-NM f_\infty} \nonumber \\
&=& \sum_{\alpha} \exp\left[-\frac{2\pi M\zeta}{N}\left\{ \left(
\Delta_{\alpha}+\bar{\Delta}_{\alpha}-\frac{c}{12}\right)\sin{\theta}
+i \left( \Delta_{\alpha}-\bar{\Delta}_{\alpha}\right)\cos{\theta}
\right\} \right] \nonumber \\
&=& (q\bar{q})^{-c/24}\ \sum_{\alpha} q^{\Delta_\alpha}
\bar{q}^{\bar{\Delta}_\alpha}
\end{eqnarray}
where $q$, the modular parameter, is given by
\begin{equation}\label{e.qdef}
q=e^{2\pi i\tau}
\end{equation}
with
\begin{equation}
\tau = \frac{M}{N}\zeta e^{i(\pi-\theta)}\ \ \ ,
\end{equation}
$\bar{q}$ is the complex conjugate of $q$ and the sum is over the
infinite set of levels whose energy $E_\alpha$ scales as
Eq.~(\ref{e.fss}). In the first part of this section we use the
notation $q$ to denote the modular parameter (Eq.~(\ref{e.qdef})).
This is not to be confused with the domain wall density.

For the Gaussian model compactified on a circle, or equivalently,
the symmetric six--vertex model in the continuum limit, the  TPF
under periodic boundary conditions in both directions is given by
the $c=1$ Coulombic partition function~\cite{Gina}
\begin{equation}\label{e.zcl}
\widetilde{{\cal Z}}_C(g)=\frac{1}{|\eta(q)|^2}\sum_{n,m\in{\bf Z}}
q^{\Delta_{n,m}(g)} \bar{q}^{\Delta_{n,-m}(g)}
\end{equation}
where $g$ is the so called Gaussian coupling constant,
$\displaystyle{ \Delta_{n,m}=\frac{1}{4} \left(\frac{n}{\sqrt{g}}+
\sqrt{g} m \right)^2}$ and $\eta(q)$ is the Dedekind eta function;
\begin{equation}
\eta(q) = q^{\frac{1}{24}} \prod_{n=1}^\infty ( 1 - q^n)\ \ \ .
\end{equation}

One can impose $U(1)$ boundary conditions on the six--vertex model
instead of periodic boundary conditions. In the Pauli spin
representation, the twisted boundary condition is
\begin{equation}
\left( \sigma_{N+1}^x \pm i\sigma_{N+1}^y \right) = e^{\mp i\varphi}
\left( \sigma_{1}^x \pm i\sigma_{1}^y \right)
\end{equation}
where $\varphi$ is the twisting angle.  The Coulombic toroidal
partition function is then modified to~\cite{Choi}
\begin{equation}\label{e.zcltw}
\widetilde{{\cal Z}}_C(g) = \frac{1}{|\eta(q)|^2} \sum_{n,m \in
{\bf Z}} e^{-in\varphi'} q^{\Delta_{n,m-\varphi/2\pi}(g)}
\bar{q}^{\Delta_{n, -(m-\varphi/2\pi)}(g)}
\end{equation}
where $\varphi$ and $\varphi'$ are the twisting angles in the space
and time directions, respectively.

After this short review, we now turn to the critical properties of
the IC phase. It is generally known that the striped IC phase is
critical and described by the $c=1$ conformal field theory in the
continuum limit~\cite{Bak,BloETAL}. In the fermion model approach,
Park and Widom~\cite{ParWid} calculated exact toroidal partition
function explicitly for the free fermion, {\it i.e.}
non-interacting domain wall system and
showed that it is of the form of Eq.~(\ref{e.zcltw}) where $g=1/2$,
$\varphi'=0$ and $\varphi/2\pi$ is the number of the domain walls
per row (mod 1). Note that the twisted boundary condition used in
Ref.~\cite{ParWid} has no direct physical meaning.

For the $T=0$ TAFIM without the second neighbor interaction, the
central charge and the scaling dimensions of several operators are
calculated analytically~\cite{BloETAL}. Since all the transfer
matrix spectra are known from the Onsager solution in this case,
one may go one step further and calculate the toroidal partition
function explicitly.  We present the calculation in
Appendix~\ref{appB}.

When $\Delta=0$, the TPF $\widetilde{\cal Z}_{\mbox{{\tiny 5-v}}}$
of the 5--vertex model can be obtained from the
$\widetilde{\cal Z}_{\mbox{{\tiny TAFIM}}}^{(\mu,\nu)}$ of
Appendix~\ref{appB}, by using the relation Eq.~({\ref{e.tafim.5v}).
The result is
\begin{eqnarray}\label{e.rt}
\widetilde{\cal Z}_{\mbox{{\tiny 5-v}}} &=&
\frac{|\tilde{q}|^{\alpha^2}}{2|\eta(\tilde{q})|^2} \left\{
|\vartheta_1(z,\tilde{q})|^2+|\vartheta_2(z,\tilde{q})|^2
+|\vartheta_3(z,\tilde{q})|^2+|\vartheta_4(z,\tilde{q})|^2 \right\}
\nonumber \\
&=& \frac{1}{|\eta(\tilde{q})|^2}\sum_{n,m\in{\bf Z}}
e^{-2i\alpha_0Mm} \tilde{q}^{(m+\frac{n+2\alpha}{2})^2/2}
\bar{\tilde{q}}^{(m-\frac{n+2\alpha}{2})^2/2} \ \ \ (\Delta=0)\ .
\end{eqnarray}
This takes the final form after the modular transformation
$\tilde{\tau} \rightarrow \tau=-1/\tilde{\tau}$;
\begin{equation}\label{e.tz5v}
\widetilde{\cal Z}_{\mbox{{\tiny 5-v}}}=\frac{1}{|\eta(q)|^2}
\sum_{m,n\in {\bf Z}} e^{-i\pi Q_1m} q^{\Delta_{m,n-Q_0}(g=1/2)}
\bar{q}^{\Delta_{m,-(n-Q_0)}(g=1/2)} \ \ \ (\Delta=0)
\end{equation}
where $Q_0$ and $Q_1$ are given in Eq.~(\ref{e.Q0}) and (\ref{e.Q1}),
respectively. This is the exactly Coulombic partition function with
the twisting angle $\varphi=2\pi Q_0$ and $\varphi'=\pi Q_1$.

Note that this can be also obtained by replacing $(m,n)$ in
Eq.~{(\ref{e.tztafim})} by $(2m,n/2)$. In Sec.~\ref{sec2}, we
gave the relation between the transfer matrices of the $T=0$ TAFIM
and the 5--vertex model. ${\bf T}_{\mbox{{\tiny 5-v}}}$ contains
odd $Q$ sectors while ${\bf T}_{\mbox{{\tiny TAFIM}}}^{(0)}$
contains the spin reversal odd sector, ${\bf T}_L-{\bf T}_R$.
So, one expects $\widetilde{\cal Z}_{\mbox{{\tiny 5-v}}}$ can be
obtained from $\widetilde{\cal Z}_{\mbox{{\tiny TAFIM}}}^{(0,0)}$
by adding terms coming from the odd $Q$ sectors and eliminating
the terms originated from ${\bf T}_L -{\bf T}_R$.
Our results show that this is exactly done by a simple substitution
of $(m,n)$ by $(2m,n/2)$.

Eq.~(\ref{e.tz5v}) implies that the IC phase of the non-interacting
domain wall model is in the universality class of the Gaussian model
with coupling constant $g=1/2$ regardless of the anisotropies in the
fugacity of the domain walls.
This result is in accord with previous works but it confirms the
universality in the strongest sense.

We assume that the effect of domain wall interactions preserves the
$c=1$ nature throughout the IC phase even though it may change the
modular parameters, the coupling constant {\it etc.} Since the
coupling constant $g$ determines the critical exponents, its
possible dependence on interactions over the IC phase is of
interest. If we denote the eigenvalue of
${\bf T}_{\mbox{{\tiny 5-v}}}$ corresponding to the $m$-th
spin wave operator in the sector $Q$ by
$e^{-E_{m,Q}}$, it is expected to take the form in the IC phase
\begin{equation}\label{e.escale}
\mbox{Re} \{E_{m,Q}\} = \frac{2\pi\zeta\sin{\theta}}{N}
\left( \frac{g}{2} (Q-Q_0)^2 + \frac{m^2}{2g}-\frac{c}{12} \right) +
Nf_{\infty}
\end{equation}
where $Q_0=qN$ is the average number of domain walls per row. Here
and below, $q$ denotes the domain wall density.
We now calculate $g$ perturbatively in the small $\Delta$ limit and
numerically for a wide range of $\Delta$.
During the perturbative calculation with $|\Delta|<1$, we will only
consider the isotropic case~$(a=1)$ for simplicity. In this case,
the eigenvalue $e^{-E_{m,Q}}$ of the transfer matrix  with
$\Delta=0$ is
\begin{equation}
\mbox{Re}\{ E_{m,Q}(\Delta=0)\} = \frac{2\pi\zeta^0}{N} \left(
\frac{m^2}{2g^0} + \frac{g^0 (Q-Q_0)^2}{2} -
\frac{c}{12}\right)+Nf_{\infty}
\end{equation}
where $c=1$, $\zeta^0 = \frac{1}{2}\tan{(\pi q/2)}$ and $g^0 =
\frac{1}{2}$ as given in Appendix~\ref{appB} and the superscripts
in $g^0$ and $\zeta^0$ denote the value for non-interacting case.
If we insert $p_j= n_j+u_j$ into the BAE where $n_j=I_j+m$ is the
solution of the $\Delta = 0$ {\em BAE} for the $m$-th excited state
in a given $Q$ sector, the resulting equation for $u_j$ is, up to
the first order in $\Delta$,
\begin{equation}
u_j = iq\left[s+\Delta e^{-qs}e^{in_j}\right]
\end{equation}
where $s$ is
\begin{equation}
s = -\Delta \frac{\sin{\pi q}}{\pi q} e^{2\pi i m/N}
\left(1+\frac{\pi^2}{6N^2}
\right)
\end{equation}
that is determined from the condition $\sum_j u_j=0$.
With this solution $\{u_j\}$, we can calculate the energy shift
$\delta E_{m,Q}\equiv E_{m,Q}(\Delta\neq 0)-E_{m,Q}(\Delta=0)$ due
to the interaction;
\begin{eqnarray}
\frac{1}{N}\mbox{Re}\left\{\delta E_{m,Q}\right\}
&=& -\sum_j \ln{\frac{ 1+e^{i(n_j+u_j)}}{1+e^{in_j}}}
= \frac{q}{N} \sum_j \frac{ se^{in_j}+ \Delta e^{2in_j}}{1+e^{in_j}}
\nonumber
\\
&=& \Delta\left(\frac{q^2}{2}+\frac{q}{2\pi}\sin{\pi q}\right) +
\Delta \left( 2\sin^2{\frac{\pi q}{2}}-\pi q \sin{\pi q}\right)
\left(\frac{m}{N}\right)^2 \nonumber \\
&&+\Delta \frac{\pi}{6N^2} \left(\frac{q}{2}\sin{\pi q}\right) +
{\cal O}\left(\frac{1}{\Delta^2}\right)\ \ \ .
\end{eqnarray}
Using the value of $\mbox{Re}\{E_{m,Q}\}/N$ at $\Delta=0$, we can
write down the
energy $\mbox{Re}\{E_{m,Q}(\Delta)\}$ up to the first order in
$\Delta$.
\begin{eqnarray}
\frac{1}{N}\mbox{Re}\{E_{m,Q}(\Delta)\}
&=& \frac{1}{N} \mbox{Re}\{E_{m,Q}(\Delta=0)\}+
\frac{1}{N}\mbox{Re}\{\delta E_{m,Q}\} \nonumber \\
&=& f'_\infty + 2\pi \left(\frac{Q-Q_0}{N}\right)^2 \left[
\frac{\zeta^0 g^0}{2}-\frac{\Delta}{8\pi}\left(\pi q\sin{\pi q}+
4\sin^2{\frac{\pi q}{2}}\right) \right] \nonumber \\
& & + 2\pi \left(\frac{m}{N}\right)^2\left[ \frac{\zeta^0}{2g^0} +
\frac{\Delta}{2\pi}\left( 2\sin^2{\frac{\pi q}{2}}-\pi q\sin{\pi q}
\right) \right] \nonumber \\
& & - \frac{\pi}{6N^2}\left(\zeta^0c^0-\frac{\Delta q}{2}\sin{\pi q}
\right) + {\cal O}\left( \frac{1}{\Delta^2} \right)
\nonumber \\
&\equiv & f'_\infty + \frac{2\pi\zeta}{N^2}\left(
\frac{g}{2}(Q-Q_0)^2 + \frac{1}{2g}m^2 - c / 12 \right) \ \ \ .
\end{eqnarray}
The new anisotropy factor $\zeta$, the Gaussian coupling constant
$g$ and the central charge $c$ are obtained by the comparing the
last two expressions;
\begin{equation}\label{e.gptb}
\left\{
\begin{array}{ccl}
c&=&1 \\
g&=& \frac{1}{2} \left(1-\frac{2\Delta}{\pi}\sin{\pi q}\right) +
{\cal O}(\Delta^2)  \\
\zeta&=& \frac{1}{2} \tan{\frac{\pi q}{2}}\left(1-2\Delta q
\cos^2{\frac{\pi q}{2}}\right) + {\cal O}(\Delta^2)\ \ \ .
\end{array}
\right.
\end{equation}
The result from the first order perturbation calculation shows that
the interaction between domain walls causes a continuous variation
of the coupling constant $g$ so the scaling dimensions vary
continuously as a function of the interaction parameter $\Delta$.

For larger values of $\Delta$, $g$ can be evaluated numerically by
the finite size corrections of the eigenvalue of the transfer
matrix (Eq.~(\ref{e.fss})). Suppose the model parameters
are tuned in such a way that $Q_0=Nq$ is an integer. That is, we are
considering the case of $q$ being integer multiple of $1/N$.
 From Eq.~(\ref{e.escale}), $g$ and $\zeta\sin\theta$ can be
evaluated if we calculate four eigenvalues
$E_{m,Q}$ with $(m,Q)=(0,Q_0),(0,Q_0{\pm}1)$ and $(1,Q_0)$.
\begin{eqnarray}
g &=& \sqrt{ {\rm Re}\{E_{0,Q_0+1}+E_{0,Q_0-1}-2E_{0,Q_0} \} /
{\rm Re}\{  E_{1,Q_0}-E_{0,Q_0} \}/2 } \nonumber \\
\zeta\sin\theta &=& \frac{N}{\pi} \sqrt{{\rm Re}
\{E_{0,Q_0+1}+E_{0,Q_0-1}-2E_{0,Q_0}\}
{\rm Re}\{E_{1,Q_0}-E_{0,Q_0}\}/2}\ \ \ .
\end{eqnarray}
Necessary $E_{m,Q}$'s are obtained by solving the BAE for $N$ up to
150. The coupling constant $g$ obtained in this way is shown in
Fig.~10 as a function of $q$ for several values of $\Delta$ for a
particular value of $a=|\Delta|+0.1$.
Note that the value of $g$ starts from around $1/2$ at $q=0$ and
ends at $1/2$ at $q=1$ and varies smoothly when $\Delta > -4$.
The values $g=1/2$ at $q=0$ is easily understood since the
interaction effect will vanish in these limit. So is the case for
$q=1$ and $\Delta+a<1$. When $\Delta\leq -4$,
the value of $g$ approaches $2$ as $q\rightarrow 1/2$.
The fact that $g=2$ exactly in the $q\rightarrow 1/2$ limit can be
derived analytically following the procedure similar to that used
by Gwa and Spohn~\cite{GwaSpo}.

The domain wall model is obtained from the TAFIM by neglecting
the spin configurations in which three spins on any elementary
triangle are in the same state. This excitation driven by the
thermal fluctuation creates or annihilates two domain walls at a
time and causes a domain wall density dislocation. (See Fig.~11.)
When two dislocations of up-triangle and down-triangle occur in
pair, the density dislocation remains as a local defect. These pair
excitations are analogous to the
vortex and anti-vortex pair excitations in the XY model.
The scaling dimension for the density dislocation~\cite{NieHil} is
$x_{0,2}\equiv \Delta_{0,2}+\Delta_{0,-2}$ since such excitation
creates or annihilates 2 domain walls.  Since $x_{0,2} = 2g$
we see that when $g<1\ (x_{0,2}<2)$ the density dislocation is
relevant and destroys the criticality of the IC phase. Therefore, if
dislocations are allowed with finite cost of energy, the IC phase
cannot remain critical and becomes the disordered fluid phase.
On the other hand,  when $g>1\ (x_{0,2}>2)$ the density dislocation
is irrelevant and the criticality of the IC phase
survives. At the boundary $g=1$, the KT transition would occur.
Since the non-interacting domain wall system has $g=1/2$, the
critical IC phase cannot survive from the density dislocation.
However as seen in Fig.~10, $g$ crosses the critical value $g=1$ in
the region of repulsive ($\Delta<0$) interactions. The Dotted line
in Fig.~7(b) inside the IC phase denotes the position where $g$
takes the value $1$.

So, we conclude that the IC phase near the $q=1/2$ C phase is stable
under the density dislocation. This shows that there are three
phases encountered if we consider the dislocation effect. They are
long-range ordered $q=1/2$ C phase, quasi-long range ordered IC
phase and the disordered phase. They are separated by the PT
transition and the KT transition. It also explains the phase
diagram of the TAFIM with the isotropic nnn interaction obtained
by Monte Carlo simulation~\cite{Lan}.

\section{Summary and Discussion}\label{sec5}
In this work, we have introduced a solvable interacting domain wall
model derived from the $T=0$ TAFIM with anisotropic nearest
neighbor and nnn interactions. The model is shown to be equivalent
to the 5--vertex model and exact phase diagram is obtained in the
three dimensional parameter space. It shows C phases where the
domain wall density is $0$, $1/2$ or $1$ and  the IC phase in
between.

The IC phase is a critical phase described by the
Gaussian fixed point. The Gaussian coupling constant $g$ which
determines the scaling dimensions of operators is a function of the
model parameters and changes smoothly from $1/2$ at $q=0$ and $q=1$
phase boundaries to $2$ at the $q=1/2$ phase boundary.
As the interaction is turned on, it decreases (increase) for
the attractive (repulsive) interaction.
For strong repulsive interactions, there is a region with $g>1$ in
which dislocation is irrelevant.
Therefore the scenario proposed by Nienhuis {\it et al.} for the
effect of the isotropic nnn interaction in the $T=0$ TAFIM is partly
born out in this model.

We also have shown by the explicit calculation of the TPF of the
non-interacting $T$=0 TAFIM that it renormalizes to the Gaussian
fixed point with the coupling constant $g$=2. This is in accord with
previous works. But, the transfer matrix spectra of the 5--vertex
model with $\Delta=0$ and that of the non-interacting $T$=0 TAFIM
are different in that some sectors
present in one are absent in the other. This re-distribution of
sectors or operator content changes $g$ of the 5--vertex model to
$1/2$ when $\Delta=0$.
The TPF  of the 5--vertex model is found to take the form of
the symmetric six--vertex model with the twisted boundary conditions.
Fractional part of the number of domain walls across a row and a
column determines the twisting angles of $U(1)$ boundary conditions
along the space and time directions, respectively.

The model considered in this work is rather special in that only one
type of domain  walls interact. For the fully interacting case, say
$y_1=y_2=y$, one need to rely on less accurate numerical methods.
The effects of the interaction between domain walls in both
direction on the phase diagram and the critical properties of the
IC phase are of interest and left for further works.

\acknowledgements
We thank discussions with F. Y. Wu and H. Park. This work is
supported by KOSEF through the grant to Center for Theoretical
Physics. J. D. Noh  also thanks Dawoo Foundation for its support.

\appendix
\renewcommand{\theequation}{A\arabic{equation}}
\setcounter{equation}{0}
\def \br{{\bf R}}
\def \b1{{\bf 1}}
\section{}\label{appA}
In this appendix, we discuss the Yang-Baxter equation~(YBE) of the
5--vertex model and alternative parametrization from which the
corresponding quantum chain hamiltonian is derived.

The YBE for the 5--vertex model is  given by
\begin{equation}\label{A1}
(\b1 \otimes \br)\ (\br'\otimes \b1)\ (\b1\otimes \br'') =
(\br''\otimes\b1)\ (\b1\otimes\br')\  (\br\otimes\b1)
\end{equation}
where \b1 is the $2\times 2$ unit matrix, $\otimes$ denotes the
direct product, \br\ is the $4\times 4$ matrix given by
\begin{equation}\label{A2}
\br = \left(
\begin{array}{cccc}
w_1 & 0 &0 & 0  \\
0 & w_5 & w_3 & 0 \\
0 & w_4 & w_6 & 0 \\
0 & 0 & 0 & w_2
\end{array}
\right)
\end{equation}
and finally $\br'$\ ($\br''$) is the same as $\br$ with $w_i$
replaced by $w_i'$\ ($w_i''$). When $w_3=w_3'=0$, the YBE has a
solution provided
\begin{equation}\label{A3}
\Delta = \frac{w_1w_2-w_5w_6}{w_2w_4} =
\frac{w_1'w_2'-w_5'w_6'}{w_2'w_4'}.
\end{equation}
The solution under the normalization $w_2=w_2'=w_2''=1$ is
\begin{equation}\label{A4}
\begin{array}{ccl}
w_1'' & = & w_1'/w_1 \\
w_3'' &=& 0 \\
w_4'' &=& (w_1w_4'-w_4w_1')/(w_5w_6) \\
w_5'' &=& w_5'/w_5 \\
w_6'' &=& w_6'/w_6
\end{array}
\end{equation}

As a result, the transfer matrix of ${\bf T}_{\mbox{{\tiny 5-v}}}$
having three independent
parameters forms a two-parameter family of commuting matrices.
Vertex weights of the 5--vertex model used in this work is given by
Eq.~(6).  If one parametrizes them alternatively as
\begin{equation}\label{A5}
\begin{array}{ccl}
w_1 &=& e^v \\
w_2  &=&1 \\
w_3  &=&0 \\
w_4  &=&(e^v-e^u) / \Delta \\
w_5  &=& w_6 = e^{u/2}
\end{array}
\end{equation}
and similarly for $w_i'$'s, the transfer matrices ${\bf T}_{
\mbox{{\tiny 5-v}}}(u,v;\Delta)$ with
different $u$ and $v$ commute, {\it i.e.}
\begin{equation}\label{A6}
\left[\  {\bf T}_{\mbox{{\tiny 5-v}}}(u,v;\Delta)\ ,\ {\bf T}_{
\mbox{{\tiny 5-v}}}(u',v';\Delta)\ \right] =0
\end{equation}
for all $u$, $u'$,$v$, and $v'$. Eq.~(\ref{A4}) with above
parametrization becomes
\begin{equation}\label{A7}
\begin{array}{ccl}
w_1'' &=& e^{v'-v} \\
w_4'' &=& e^v( e^{v'-v} - e^{u'-u} ) / \Delta \\
w_5'' &=& w_6'' = e^{(u'-u)/2}
\end{array}
\end{equation}

Standard parametrizations of solutions to the YBE involve the
so-called spectral parameter $u$ with which the YBE displays the
difference property; {\it i.e.}, if $\br = \br(u)$ and
$\br'=\br(u')$ then $\br''=\br(u'-u)$.
At criticality, it gives the physical meaning of the anisotropy
angle~\cite{KimP}.
Also, corresponding quantum chain hamiltonian commuting with the
transfer matrix is obtained by the logarithmic derivative at $u=0$.
We find from Eq.~(\ref{A7}) that the 5--vertex model also displays
the difference property if we set $v=0$. This is the special case
$w_1=w_2$ considered in Ref.~\cite{GulBL}.

We calculated the quantum hamiltonian $\hat{{\cal H}}$
of the one-dimensional quantum spin chain by taking the logarithmic
derivative of the transfer matrix at $u=0$ for the case of $v=0$.
The result is
\begin{eqnarray}
\hat{{\cal H}}&=&\left. {\bf T}_{\mbox{{\tiny 5-v}}}^{-1}\ \
\frac{\partial {\bf T}_{\mbox{{\tiny 5-v}}}}{\partial u}
\right|_{u,v=0} \nonumber \\
&=& \sum_{i=1}^N \left\{\  \hat{s}^+_i\  \hat{s}^-_{i+1} +
\frac{\Delta}{4} \ \hat{s}^z_i\ \hat{s}^z_{i+1} \right\}
\end{eqnarray}
where $\hat{s}_i$ is the quantum spin density operator at site $i$.
This non-hermitian hamiltonian is similar to the hamiltonian of the
XXZ quantum spin chain. The difference is that there is no term
$\hat{s}^-_i \hat{s}^+_{i+1}$ in this model.
So, there is a net flux of the spin flow from the right to the left
of the chain. It comes from the anisotropic choice of the vertex
weights at the beginning.

\renewcommand{\theequation}{B\arabic{equation}}
\setcounter{equation}{0}
\section{}\label{appB}
In this appendix, we present the phase diagram of the $T=0$ TAFIM
with anisotropic nearest neighbor interaction and the toroidal
partition functions under the general boundary conditions.

Through the star-triangle relation, the Ising model on the
triangular lattice can be mapped into the Ising model on the honeycomb
lattice~\cite{Bax}.
Let $K_i=K+\delta_i$ and $L_i \ (i=1,2,3)$ be the interaction
strength (including the factor $-1/kT$) of the Ising model on the
triangular and honeycomb lattice, respectively.
Then, the partition functions ${\cal Z}_{\mbox{{\tiny TAFIM}}}$ on a
triangular lattice with ${\cal N}$ sites and
${\cal Z}_{\mbox{\tiny H}}$ on a honeycomb lattice with $2{\cal N}$
sites are related by
\begin{equation}
{\cal Z}_{\mbox{\tiny H}}(\mbox{\boldmath $L$},2{\cal N}) =
R^{\cal N} {\cal Z}_{\mbox{{\tiny TAFIM}}}(
\mbox{\boldmath $K$},{\cal N})
\end{equation}
provided {\boldmath $K$} and {\boldmath $L$} satisfy the star-triangle
relation:
\begin{equation}
\exp\left[ K_1 s_2 s_3 + K_2 s_3s_1 + K_3s_1s_2\right] =
R \sum_{t=\pm 1} \exp\left[ t(L_1s_2+L_2s_2+L_3s_3)\right] \ \ \ .
\end{equation}
Here $R$ is the normalization factor.

If we take the zero temperature limit $K
\rightarrow -\infty$, the solution of the star-triangle relation is
\begin{equation}\label{e.kl}
\sinh{2L_i}=\frac{z_i}{k}\ ,\ \ \cosh{2L_i}=\frac{z_j^2+z_k^2-z_i^2}{
2z_j z_k}\ ,\ \ R^2 = \frac{2z_1z_2z_3}{k^2}
\end{equation}
where
$$ z_i=e^{2\delta_i}\ \ \ ,$$
$$ k^2 = (4z_i^2z_j^2z_k^2)/((z_j^2+z_k^2-z_i^2)^2-4z_j^2z_k^2)$$
and $(i,j,k)$ is a cyclic permutation of (1,2,3).

Now, we consider the transfer matrix $\mbox{\bf T}^{(\mu)}_{
\mbox{\tiny H}}$ on the honeycomb lattice whose matrix element
$\mbox{\bf T}^{(\mu)}_{\mbox{\tiny H}}(
\mbox{\boldmath$s$},\mbox{\boldmath$t$})$ is the Boltzmann weight
for a spin configuration shown in Fig.~12;
\begin{eqnarray}\label{e.thc}
\mbox{\bf T}^{(\mu)}_{\mbox{\tiny H}}
(\mbox{\boldmath$s$},\mbox{\boldmath$t$}) &=&
\sum_{r_i=\pm 1}\exp\left[\sum_{m=1}^{M/2}L_3 t_{2m}t_{2m+1}\right]
\times \exp\left[ \sum_{m=1}^{M/2} ( L_1 r_{2m-1}t_{2m-1} +
L_2 r_{2m} t_{2m} \right]\nonumber \\
&& \times \exp\left[\sum_{m=1}^{M/2} L_3 r_{2m-1} r_{2m} \right]
\times\exp\left[ \sum_{m=1}^{M/2} ( L_1 s_{2m}r_{2m} + L_2 s_{2m-1}
r_{2m-1} \right] \nonumber \\
&\equiv&\left(\mbox{T}_A\cdot\mbox{T}_{B1} \cdot \mbox{T}_C \cdot
\mbox{T}_{B2} \right) (\mbox{\boldmath$s$},\mbox{\boldmath$t$})
\end{eqnarray}
The superscript $\mu\ (=0,1)$ denotes the boundary condition
$s_{M+1,j} = (-1)^{\mu} s_{1,j}$.
Each of the four factors in the first two lines in Eq.~(\ref{e.thc})
defines the four matrices $\mbox{T}_A$, $\mbox{T}_{B1}$,
$\mbox{T}_C$ and $\mbox{T}_{B2}$, respectively.
Their operator forms are
\begin{equation}
\left\{
\begin{array}{ccl}
\mbox{T}_A &=& \prod_{m=1}^{M/2}
\exp\left[L_3\sigma_{2m}^z\sigma_{2m+1}^z \right] \\
\mbox{T}_{B1} &=& \left( \frac{4z_1z_2}{k^2} \right)^{M/4}
\prod_{m=1}^{M/2} \exp\left[L_1^{*}\sigma_{2m-1}^x + L_2^{*}
\sigma_{2m}^x \right] \\
\mbox{T}_{B2} &=& \left( \frac{4z_1z_2}{k^2} \right)^{M/4}
\prod_{m=1}^{M/2} \exp\left[L_2^{*}\sigma_{2m-1}^x + L_1{*}
\sigma_{2m}^x \right] \\
\mbox{T}_{C} &=& \prod_{m=1}^{M/2}\exp\left[L_3\sigma_{2m-1}^z
\sigma_{2m}^z \right]
\end{array} \right.
\end{equation}
where $\hbox{\boldmath $\sigma$}_i^{\mu}$ is the Pauli spin operator
at site $i$ and $e^{-2L_i^{*}} = \tanh{L_i}$.
Putting $N$ rows of Fig.~12 in succession and applying the
star--triangle transformation gives the $M\times N$ triangular
lattice whose basis vectors are rotated by $90^0$ from those of
Fig.~1(a). Thus, to calculate the toroidal partition function of
$T=0$ TAFIM, we instead carry
out the calculation using ${\bf T}^{(\mu)}_{\mbox{\tiny H}}$.
The boundary condition along the vertical direction is $s_{i,N+1}
=(-1)^{\nu}s_{i,1}$.
For each boundary condition, the partition function ${\cal Z}_{
\mbox{{\tiny TAFIM}}}^{(\mu,\nu)}$ is
written as
\begin{equation}
{\cal Z}_{\mbox{{\tiny TAFIM}}}^{(\mu,\nu)} =
\mbox{Tr}\ \left[{{\bf T}^{(\mu)}_{\mbox{\tiny H}}}
\right]^N\ {\bf R}^{\nu}
\end{equation}
where ${\bf R}$ is the spin reversal operator.

Following the same procedure as in Ref.~\cite{Wan}, we diagonalized
the transfer matrix Eq.~(\ref{e.thc}) exactly. If
we write the transfer matrix as
${\bf T}^{(\mu)}_{\mbox{\tiny H}} = \left(4z_1 z_2/k^2
\right)^{M/2} e^{-\hat{\cal H}}$, then the hamiltonian
$\hat{\cal H}$ after the usual Wigner--Jordan transformation can be
written as
\begin{equation}
\hat{{\cal H}} = \frac{1}{2} \sum_{k} \left\{ \epsilon_1(k)
( 2 \hat{n}_k - 1 ) + \epsilon_2(k) (2 \hat{m}_k -1 ) \right\}
\end{equation}
where $\hat{n}_k$ and $\hat{m}_k$ are the mutually commuting
occupation number operators with eigenvalue $n_k,m_k=0,1$ and
$\epsilon_1(k)$ and $\epsilon_2(k)$ are the quasi-particle
excitation energy which is given by
\begin{eqnarray}
\epsilon_1(k) &=& \left|\mbox{sgn}(k)\cosh^{-1}{t_+}-
\cosh^{-1}{p}\right|-i \cos^{-1}{t_-} \nonumber \\
\epsilon_2(k) &=&\left|\mbox{sgn}(k)\cosh^{-1}{t_+}+\cosh^{-1}{p}
\right|-i \cos^{-1}{t_-}
\end{eqnarray}
where
\begin{eqnarray}
t_{\pm} &=& \frac{z_3^2}{4z_1z_2} \pm \frac{1}{2}
\sqrt{4+\frac{z_3^4}{4z_1^2z_2^2}-\frac{2z_3^2\cos{k}}{z_1z_2}}
\nonumber \\ p &=& \frac{z_1^2+z_2^2}{2z_1z_2} \ \ \ .
\end{eqnarray}
 From now on, we set $z_3=1$ since all quantities are functions of
$z_1/z_3$ and $z_2/z_3$ only. $z_i/z_3$ in this section is equal to
$x_i$ of the text if $\varepsilon_i=0$. (See Eq.~(4).)
The values of $k$ are restricted to the set
\begin{equation}\label{e.kval}
k = \frac{2\pi}{M} \times \left\{
\begin{array}{cl}
2{\bf Z}-1+\mu&\hspace{1cm} \mbox{if } \hspace{0.5cm}\sum_k(n_k+m_k)
= \mbox{ even} \\
2{\bf Z}+\mu&\hspace{1cm} \mbox{if } \hspace{0.5cm}\sum_k(n_k+m_k)
= \mbox{ odd} \end{array} \right.
\end{equation}
for boundary condition $\mu$ along the horizontal direction.

The partition function ${\cal Z}_{\mbox{{\tiny TAFIM}}}^{(\mu,\nu)}$
is given by
\begin{equation}\label{e.b11}
{\cal Z}_{\mbox{{\tiny TAFIM}}}^{(\mu,\nu)} = \left( e^{-N E_e^{\mu}}
{\cal Z}_{even}^{\mu}+(-1)^{\nu} e^{-N E_o^{\mu}}
{\cal Z}_{odd}^{\mu}\right)\left(\frac{2}{z_3}\right)^{-{\cal N}/2}
\end{equation}
where
\begin{equation}
{\cal Z}_{even(odd)}^{\mu}=\sum_{\mbox{\tiny even(odd)}}
\exp\left\{ -N \sum_k \left(
\epsilon_1(k)n_k + \epsilon_2(k) m_k \right) \right\} \ \ \ ,
\end{equation}
$\sum_{\mbox{\tiny even}}$ $(\sum_{\mbox{\tiny odd}})$ denoting the
sums over the occupation number configurations $\{n_k,m_k=0,1\}$
under the restriction $\sum_k(n_k+m_k)=$ even (odd), respectively,
and the values of $k$ are given in Eq.~(\ref{e.kval}). We will say a
state is in an even (odd) sector if $\sum_k(n_k+m_k)$ is even (odd).
In Eq.~(\ref{e.b11}) $E_e^{\mu}$ and $E_o^{\mu}$ are the ground
state energy in the respective sectors and are given by
\begin{eqnarray}
E_e^{\mu} &=& -\frac{1}{2} \sum_{n=1}^{M/2} \left( \epsilon_1\left[
\frac{2\pi}{M}(2n{-}1{+}\mu)\right]+\epsilon_2\left[\frac{2\pi}{M}
(2n{-}1{+}\mu)\right] \right) = -\sum_n \epsilon_1\left[
\frac{2\pi}{M}(2n{-}1{+}\mu)\right] \nonumber \\
E_o^{\mu} &=& -\frac{1}{2} \sum_{n=1}^{M/2} \left( \epsilon_1\left[
\frac{2\pi(2n{+}\mu)}{M}\right] + \epsilon_2\left[\frac{2\pi (2n{+}
\mu)}{M}\right] \right)
= -\sum_n \epsilon_1\left[\frac{2\pi (2n{+}\mu)}{M}\right] \ \ \ .
\end{eqnarray}
We obtain the finite corrections to $E_e^{\mu}$ and $E_o^{\mu}$ from
the Euler-Mclaurin formula. The results for periodic boundary
condition $(\mu=0)$ are
\begin{eqnarray}
E_e^0 &=& \left\{
\begin{array}{ll}
-\frac{M}{4\pi}\int_0^{2\pi}\epsilon_1(k)dk-\frac{4\pi\epsilon_0'}{M}
\left( \frac{1}{12} - \alpha^2 \right) & , \alpha \leq \frac{1}{2} \\
-\frac{M}{4\pi}\int_0^{2\pi}\epsilon_1(k)dk-\frac{4\pi\epsilon_0'}{M}
\left(\frac{1}{12}-(1-\alpha)^2\right) & , \alpha \geq \frac{1}{2}
\end{array} \right. \nonumber \\
E_o^0&=&-\frac{M}{4\pi}\int_0^{2\pi} \epsilon_1(k) dk -
\frac{4\pi\epsilon_0'}
{M}\left( \frac{1}{12} - (1/2-\alpha)^2 \right)
\end{eqnarray}
where
\begin{eqnarray}
\epsilon_0' &=& \left.\frac{d\epsilon_1}{dk}\right|_{k=k_c^{+}} =
\frac{\sqrt{(2z_1^2+2z_2^2-1)-(z_1^2-z_2^2)^2}}{2(z_1^2+z_2^2)-1}
\nonumber \\
\alpha &=& \frac{k_c M}{4\pi} - \left[\frac{k_c M}{4\pi}\right]
\end{eqnarray}
with $[x]$ denoting the integer part of $x$. The ground state
energies for each sector under anti-periodic boundary condition
$(\mu=1)$ are
\begin{equation}
E_{e}^1 = E_{o}^0 \ \ \ , \hspace{2cm} E_{o}^1 = E_{e}^0 \ \ \ .
\end{equation}
The quantity
$\frac{1}{2}\ln{(z_3/2)}-\frac{1}{4\pi}\int_0^{2\pi}\epsilon_1(k)dk$
is the bulk free energy $f_\infty$ per site.

 From the predictions of the conformal field theory, we know that the
transfer matrix has gapless excitations with linear dispersion
relation at the criticality.
The quasi-particle excitation energies become zero at
$k=\pm k_c$ where
\begin{equation}\label{e.kcdef}
\cos{k_c}=\frac{1}{2z_1z_2}\left[(z_1^2+z_2^2)-(z_1^2-z_2^2)^2\right]
\end{equation}
in the range $|z_1-z_2| \leq z_3$ and $|z_1+z_2| \geq z_3$. So,
we conclude that the system is critical in this range.
This includes the result of
Bl\"{o}te and Hilhorst~\cite{BloHil} who treated the case $z_1=z_2$.

The toroidal partition function
$\widetilde{\cal Z}_{\mbox{{\tiny TAFIM}}}^{(\mu,\nu)}$ under the
general boundary condition $(\mu,\nu)$ is given as
\begin{equation}
\widetilde{\cal Z}_{\mbox{{\tiny TAFIM}}}^{(\mu,\nu)} =
\lim_{\stackrel{N,M\rightarrow\infty}{M/N = fixed}}
{\cal Z}_{\mbox{{\tiny TAFIM}}}^{(\mu,\nu)} / e^{-NMf_\infty}\ \ \ .
\end{equation}
Especially, the toroidal partition function for periodic boundary
condition in both directions is given as
\begin{equation}
\widetilde{\cal Z}_{\mbox{{\tiny TAFIM}}}^{(0,0)} =
\lim_{\stackrel{N,M\rightarrow\infty}{N/M = fixed}}
e^{\frac{4\pi N}{M}\epsilon_0'(1/12-\alpha^2)} {\cal Z}_{even}^0
+e^{\frac{4\pi N}{M}\epsilon_0'(1/12-(1/2-\alpha)^2)}
{\cal Z}_{odd}^0 \ \ \ .
\end{equation}
Since $N$ is large, the modes near $k=\pm k_c$ whose energy scales
like $1/M$ contribute factors of ${\cal O}(1)$ in the sum.
Therefore, for $M,N \rightarrow \infty$ with $N/M$ fixed, we may
replace the dispersion relation by the linear one
\begin{equation}\label{e.lindsp}
\epsilon_{1,2}(k) = \epsilon_o' |k\mp k_c| - i\alpha_0'(k-k_c)\mp i
\alpha_0
\end{equation}
where
\begin{eqnarray}\label{e.alpha0def}
\alpha_0' &=&\left.\frac{d\alpha}{dk}\right|_{k=k_c^+}=\frac{z_2^2-
z_1^2}{2(z_1^2+z_2^2)-1} \nonumber \\
\cos\alpha_0&=&\cos\alpha(k_c)=\frac{1-z_1^2-z_2^2}{2z_1z_2}\ \ \ .
\end{eqnarray}
The restricted sums can be done conveniently using the transformation
\begin{eqnarray}
\sum_{\mbox{\tiny even}}&=&\left[\mbox{tr}_n^e\frac{1}{2}\left\{1+
(-1)^{\sum_k n_k}\right\}\right]\left[\mbox{tr}_m^e\frac{1}{2}
\left\{1+ (-1)^{\sum_k m_k}\right\}\right] \nonumber \\
&+& \left[ \mbox{tr}_n^e \frac{1}{2} \left\{1-(-1)^{\sum_k n_k}
\right\}\right] \left[\mbox{tr}_m^e\frac{1}{2}\left\{1-(-1)^{
\sum_km_k}\right\}\right] \nonumber \\
\sum_{\mbox{\tiny odd}}  &=& \left[ \mbox{tr}_n^o \frac{1}{2}
\left\{1+ (-1)^{\sum_k n_k}\right\}\right]\left[\mbox{tr}_m^o
\frac{1}{2}\left\{1-(-1)^{\sum_k m_k}\right\}\right] \nonumber \\
&+& \left[ \mbox{tr}_n^o \frac{1}{2} \left\{1-(-1)^{\sum_k n_k}
\right\}\right] \left[\mbox{tr}_m^o\frac{1}{2}\left\{1+(-1)^{
\sum_km_k}\right\}\right]
\end{eqnarray}
where
$$
\mbox{tr}_n^e = \prod_{k=\frac{2\pi}{M}(2n-1)} \sum_{n_k=0,1}\ \ \
\mbox{and } \ \  \ \ \ \
\mbox{tr}_n^o = \prod_{k=\frac{4\pi n}{M}} \sum_{n_k=0,1}
$$
and similarly for $\mbox{tr}_m^e$ and $\mbox{tr}_m^o$.

After a lengthy calculation with this linear dispersion relation
Eq.~(\ref{e.lindsp}), we obtain
\begin{equation}
\widetilde{\cal Z}_{\mbox{{\tiny TAFIM}}}^{(0,0)} =
\frac{|\tilde{q}|^{\alpha^2}}{2|\eta(\tilde{q})|^2} \left\{
-|\vartheta_1(z,\tilde{q})|^2+|\vartheta_2(z,\tilde{q})|^2
+|\vartheta_3(z,\tilde{q})|^2+|\vartheta_4(z,\tilde{q})|^2 \right\}
\end{equation}
where $\eta(q) = q^{1/24} \prod_{n=1}^{\infty}(1-q^n)$ is the
Dedekind eta function, $\vartheta_i$ are the  Jacobi theta
function~\cite{Gina}, $\tilde{q}=e^{2\pi i\tilde{\tau}}$ with
$$
\tilde{\tau} = \frac{2i N}{M} (\epsilon_0'+i\alpha_0') \\
$$
and finally
$$
z = \frac{\alpha_0N}{2} + \pi\alpha\tilde{\tau}\ \ \ .
$$

Following the same procedures, we can also calculate
$\widetilde{\cal Z}_{\mbox{{\tiny TAFIM}}}^{(\mu,\nu)}$
under the general boundary condition $(\mu,\nu)$. We present only
the results.
\begin{eqnarray}
\widetilde{\cal Z}_{\mbox{{\tiny TAFIM}}}^{(1,0)} &=&
\frac{|\tilde{q}|^{\alpha^2}}{2|\eta(\tilde{q})|^2} \left\{
|\vartheta_1(z,\tilde{q})|^2+|\vartheta_2(z,\tilde{q})|^2
+|\vartheta_3(z,\tilde{q})|^2-|\vartheta_4(z,\tilde{q})|^2 \right\}
\nonumber \\
\widetilde{\cal Z}_{\mbox{{\tiny TAFIM}}}^{(0,1)} &=&
\frac{|\tilde{q}|^{\alpha^2}}{2|\eta(\tilde{q})|^2} \left\{
|\vartheta_1(z,\tilde{q})|^2-|\vartheta_2(z,\tilde{q})|^2
+|\vartheta_3(z,\tilde{q})|^2+|\vartheta_4(z,\tilde{q})|^2 \right\}
\nonumber \\
\widetilde{\cal Z}_{\mbox{{\tiny TAFIM}}}^{(1,1)} &=&
\frac{|\tilde{q}|^{\alpha^2}}{2|\eta(\tilde{q})|^2} \left\{
|\vartheta_1(z,\tilde{q})|^2+|\vartheta_2(z,\tilde{q})|^2
-|\vartheta_3(z,\tilde{q})|^2+|\vartheta_4(z,\tilde{q})|^2 \right\}
\end{eqnarray}

The toroidal partition function $\widetilde{\cal Z}_{\mbox{
{\tiny TAFIM}}}^{(\mu,\nu)}$ can be rewritten as
an infinite series in $\tilde{q}$. Using the series form of
$\vartheta_i$'s and rearranging the summands, we obtain
\begin{equation}\label{e.tztimn}
\widetilde{\cal Z}_{\mbox{{\tiny TAFIM}}}^{(\mu,\nu)}=
\frac{1}{|\eta(\tilde{q})|^2}\sum_{n,m\in{\bf Z}}
(-1)^{\nu}e^{i\alpha_0Nm}
\tilde{q}^{((2n+\mu)/2+\alpha+m/2)^2/2}\
\bar{\tilde{q}}^{((2n+\mu)/2+\alpha-m/2)^2/2}
\end{equation}

To compare with the triangular lattice shown in Fig.~1(a), we
perform the modular transformation $\tilde{\tau}\rightarrow
\tau = -1/\tilde{\tau}$. This is
achieved by applying Poisson sum formula to both summation indices
$n$ and $m$ in Eq.~(\ref{e.tztimn}). The resulting expression for
the periodic boundary condition $(\mu=0,\nu=0)$ is
\begin{equation}\label{aaa}
\widetilde{\cal Z}_{\mbox{{\tiny TAFIM}}}^{(0,0)} =
\frac{1}{|\eta(q)|^2} \sum_{m,n\in {\bf Z}} e^{-2\pi i\alpha m}
q^{(m/2+(n-\alpha_0 N/2\pi))^2/2}
\bar{q}^{(m/2-(n-\alpha_0 N/2\pi))^2/2}
\end{equation}
where $q=e^{2\pi i\tau}$. Note that
$$\tau = \frac{M}{N}\zeta^0 e^{i(\pi-\theta_0)}$$
with
$$\zeta^0=\left( 2(z_1^2+z_2^2)-1\right)^{1/2}/2$$
and
$$\pi-\theta_0 =
\cos^{-1}{\frac{z_2^2-z_1^2}{\sqrt{2(z_1^2+z_2^2)-1}}}.$$
This is exactly the Coulombic partition function with the twisted
boundary conditions and the coupling constant $g=2$.

Since $z_i$'s are the activities of the diamonds shown in Fig.~1(b),
one easily obtains~\cite{BloHil} from the bulk free energy that the
mean domain wall densities of each type are given by
\begin{eqnarray}
\langle n_1+n_2 \rangle / {\cal N} = \alpha_0 / \pi \nonumber \\
\langle n_1-n_2 \rangle / {\cal N} = k_c / \pi
\end{eqnarray}
with $\alpha_0$ and $k_c$ are given by Eq.~(\ref{e.alpha0def}) and
(\ref{e.kcdef}), respectively. Therefore, one sees that the twisting
angles are $\alpha_0 N$ and $-2\pi \alpha$ in Eq.~(\ref{aaa}) are
related to the total domain wall densities as
\begin{equation}\label{e.Q0}
\alpha_0 N = \frac{\pi\langle n_1+n_2 \rangle }{M} \equiv {\pi Q_0}
\end{equation}
and
\begin{equation}\label{e.Q1}
2\pi \alpha = \frac{\pi\langle n_1-n_2 \rangle }{N}\equiv
\frac{\pi Q_1}{2}
\end{equation}
where $Q_0$ is the number of domain walls per row and $Q_1$ is the
difference of number of type 1 domain walls and that of type 2 per
column.  Using this quantities, $\widetilde{\cal Z}_{
\mbox{{\tiny TAFIM}}}^{(0,0)}$ is written as
\begin{equation}\label{e.tztafim}
\widetilde{\cal Z}_{\mbox{{\tiny TAFIM}}}^{(0,0)} =
\frac{1}{|\eta(q)|^2} \sum_{m,n\in {\bf Z}} e^{-i\pi Q_1 m/2}
q^{\Delta_{m,n-Q_0/2}(g=2)}\bar{q}^{\Delta_{m,-(n-Q_0/2)}(g=2)}
\end{equation}
where $\Delta_{m,n}(g)=\left( m/\sqrt{g}+\sqrt{g}n\right)^2/4$.

\pagebreak
\noindent
{\Large \bf Figure Captions}

\noindent
Fig. 1. {(a) Correspondence between the anisotropic couplings and
the lattice directions in the TAFIM with nearest and next nearest
neighbor interactions. (b) Labeling of three types of diamonds.
Types 1 and 2 are considered as domain wall excitations.}

\noindent
Fig. 2. {A typical striped domain wall configuration derived from a
ground state of the TAFIM on $4\times 6$ lattice. The filled circles
represent spin up states and empty circles represent spin down
states. We use the periodic boundary condition along the horizontal
direction and the anti-periodic boundary condition along the
vertical direction. So, the resulting domain wall configuration has
even $Q$ and odd $\Omega$.}

\noindent
Fig. 3. {Four possibilities of next nearest neighbor bonds along the
direction 1.}

\noindent
Fig. 4. {Deformation of Fig. 2 into a square lattice.}

\noindent
Fig. 5. {Five types of unit squares in the deformed lattice and
assignment of vertex configurations.}

\noindent
Fig. 6. {Typical $x_2$ dependence of $q$ within the IC phase is shown
for $x_1=x_2$. The curves are for $\Delta=-0.7,-0.3,0,0.3$ and $0.7$,
respectively.}

\noindent
Fig. 7. {(a) Phase diagram in $\ln{x_2}$--$\Delta$ plane for $a=1$.
$\langle 0\rangle$, $\langle 1\rangle$ and $\langle$IC$\rangle$
denotes the C phase with $q=0,1$ and the IC phase, respectively.
(b) Same as in (a) with $a=7$. New C phase with $q=1/2$ appears for
$\Delta<-4$.} The Dotted line in the IC phase denotes the position
where dislocations become irrelevant. See section \ref{sec4}.

\noindent
Fig. 8. {A typical root distribution of the BAE in complex $\alpha$
plane.  This figure shows the solutions of BAE for $q=1/2$ and
$\Delta=-7$.}

\noindent
Fig. 9. {Typical $x_2$ dependence of The domain wall densities for
$\Delta=-4.0$ and $-5.0$. The curves are for  $a=\Delta+0.1$. This is
obtained from the equation of state ( Eq.~(\ref{e.eos})) where
configurational free energy is taken from the numerical solution of
the BAE with $N=150$.}

\noindent
Fig. 10.  {The Gaussian coupling constant $g$ calculated numerically
with lattice size $N=150$ and $\Delta=0.5,0.0,-0.5,-2.0,-3.0,-4.0$
and $-5.0$. For each curve, $a$ is set to the value
$a=|\Delta|+0.1$.}

\noindent
Fig. 11.  {Effect of finite temperature in TAFIM is to excite
dislocations in domain walls. Simultaneous creation of vortex (a)
and anti-vortex (b) pairs as in (c) distroys the IC phase if $g<1$
but is irrelevant if $g>1$.}

\noindent
Fig. 12. {The honeycomb lattice transfer matrix.}

\end{document}